\documentclass[12pt]{article}
\usepackage{amsmath,amssymb,amsfonts,color}
\usepackage{graphicx,psfrag,epsf}
\usepackage{enumerate}

\usepackage{url} % not crucial - just used below for the URL 

\usepackage{graphicx,rotating,booktabs}

\usepackage{geometry}
\RequirePackage{amsthm,amsmath,amsfonts,amssymb}
\RequirePackage[authoryear,round]{natbib}
\RequirePackage[colorlinks,citecolor=blue,urlcolor=blue]{hyperref}
\RequirePackage{graphicx}
\RequirePackage{bbm}
\RequirePackage{mathtools}
\RequirePackage{algorithm}
\RequirePackage{algpseudocode}
\RequirePackage{geometry}
\RequirePackage{booktabs}
\RequirePackage{multirow}

\def\tp{{\mathsf{T}}}

\newcommand{\appropto}{\mathrel{\vcenter{\offinterlineskip\halign{\hfil$##$\cr\propto\cr\noalign{\kern2pt}\sim\cr\noalign{\kern-2pt}}}}}
\newcommand{\CC}{C\nolinebreak\hspace{-.05em}\raisebox{.4ex}{\tiny\bf+}\nolinebreak\hspace{-.10em}\raisebox{.4ex}{\tiny\bf +}}
\DeclareMathAlphabet\mathbfcal{OMS}{cmsy}{b}{n}

\DeclareMathOperator*{\argmin}{arg\,min}
\DeclareMathOperator{\Tr}{Tr}
\DeclarePairedDelimiter\abs{\lvert}{\rvert}

\makeatletter
\let\oldabs\abs
\def\abs{\@ifstar{\oldabs}{\oldabs*}}

\algblock{ParFor}{EndParFor}
\algnewcommand\algorithmicparfor{\textbf{parallel for}}
\algnewcommand\algorithmicpardo{\textbf{do}}
\algnewcommand\algorithmicendparfor{\textbf{end\ parallel for}}
\algrenewtext{ParFor}[1]{\algorithmicparfor\ #1\ \algorithmicpardo}
\algrenewtext{EndParFor}{\algorithmicendparfor}

\def\vectorfontone{\bf}
\def\vzero{{\vectorfontone 0}}

\def\vd{{\vectorfontone d}}
\def\ve{{\vectorfontone e}}

\def\vI{{\vectorfontone I}}
\def\vm{{\vectorfontone m}}
\def\vr{{\vectorfontone r}}

\def\vu{{\vectorfontone u}}
\def\vv{{\vectorfontone v}}

\def\vx{{\vectorfontone x}}

\def\vy{{\vectorfontone y}}

\def\vz{{\vectorfontone z}}

\def\vectorfonttwo{\boldsymbol}
\def\valpha{{\vectorfonttwo \alpha}}
\def\vbeta{{\vectorfonttwo \beta}}

\def\vvarepsilon{{\vectorfonttwo \varepsilon}}
\def\vgamma{{\vectorfonttwo \gamma}}

\def\vtheta{{\vectorfonttwo \theta}}

\def\vmu{{\vectorfonttwo \mu}}

\def\matrixfontone{\bf}
\def\mzero{{\matrixfontone 0}}

\def\mA{{\matrixfontone A}}
\def\mB{{\matrixfontone B}}

\def\mE{{\matrixfontone E}}

\def\mI{{\matrixfontone I}}

\def\mL{{\matrixfontone L}}
\def\mM{{\matrixfontone M}}

\def\mQ{{\matrixfontone Q}}

\def\mS{{\matrixfontone S}}
\def\mT{{\matrixfontone T}}

\def\matrixfonttwo{\boldsymbol}
\def\mSigma{{\matrixfonttwo \Sigma}}

\def\mOmega{{\matrixfonttwo \Omega}}

\def\mPsi{{\matrixfonttwo \Psi}}

\def\sD{{\mathcal D}}

\title{Scalable Expectation Propagation for Mixed-Effects Regression}
\author{Jackson Zhou, John T. Ormerod, and Clara Grazian}
\date{}

\begin{document}

\maketitle

\begin{abstract}
Mixed-effects regression models represent a useful subclass of regression models for grouped data; the introduction of random effects allows for the correlation between observations within each group to be conveniently captured when inferring the fixed effects.
At a time where such regression models are being fit to increasingly large datasets with many groups, it is ideal if (a) the time it takes to make the inferences scales linearly with the number of groups and (b) the inference workload can be distributed across multiple computational nodes in a numerically stable way, if the dataset cannot be stored in one location.
Current Bayesian inference approaches for mixed-effects regression models do not seem to account for both challenges simultaneously.
To address this, we develop an expectation propagation (EP) framework in this setting that is both scalable and numerically stable when distributed for the case where there is only one grouping factor.
The main technical innovations lie in the sparse reparameterisation of the EP algorithm, and a moment propagation (MP) based refinement for multivariate random effect factor approximations.
Experiments are conducted to show that this EP framework achieves linear scaling, while having comparable accuracy to other scalable approximate Bayesian inference (ABI) approaches.
\end{abstract}

\section{Introduction}\label{sec:intro}

Grouped data are a common occurrence in many statistical domains.
When inferring effects from such data, mixed-effects regression models (such as generalised linear mixed models) have been widely used to account for correlated observations within each group; \cite{pinheiro2006mixed}, \cite{mcculloch2011generalized}, and \cite{faraway2016extending} are good introductory texts in this regard.
In this mixed-effects regression setting, it is often the case that a Bayesian approach is preferred, in order to regularise the likelihood and avoid computational difficulties \citep{faraway2016extending}, and/or to incorporate prior information into the parameter estimation process.

Markov chain Monte Carlo (MCMC) methods such as Hamiltonian Monte Carlo (HMC, \citealt{mackay2003information}) have traditionally been used to perform inference on the Bayesian posterior distribution. This is helped in no small part by the accessibility and ease-of-implementation of MCMC programs such as JAGS \citep{plummer2003jags} and Stan \citep{carpenter2017stan}.
In addition to MCMC, popular alternatives lie in ABI methods, which tend to be less accurate but are orders of magnitude faster.
Popular ABI methods include variational Bayes (VB) methods \citep{blei2017variational}, EP \citep{minka2001family}, and integrated nested Laplace approximations (INLA, \citealt{rue2009approximate}).

Nowadays, it is not uncommon to use a mixed-effects modelling framework for large datasets with many groups; this can be seen in fields such as epidemiology (e.g., \citealt{ahn2012modeling}), genomics (e.g., \citealt{zhou2020scalable}), and recommendation modelling (e.g., \citealt{zhang2016glmix}).
Unfortunately, the previously described Bayesian inference methods present some unique challenges in this setting.
Importantly, we believe that these challenges have only been partially addressed in the current literature.

The first challenge is scalability with respect to the number of groups in the data.
Ideally, the run time of the inference method scales linearly (or better) with respect to the number of groups so that inference is feasible for large datasets with many groups.
This is equivalent to the method scaling linearly with respect to the number of random effects, as the number of groups is generally proportional to the number of random effects.
Such a time complexity is typically not possible for standard MCMC methods. For example, \cite{neal2012mcmc} demonstrates that HMC can be expected to scale according to $d^{5/4}$, where $d$ is the number of model parameters (these include the random effects).
Similarly, for an off-the-shelf full-rank Gaussian variational approximation such as the automatic differentiation variational inference approach of \cite{kucukelbir2017automatic}, the run time scales quadratically with respect to the number of model parameters \citep{tan2018gaussian}.
There do exist approaches which scale linearly however. Perhaps the most notable example is the integrated nested Laplace approximations (INLA) methodology of \cite{rue2009approximate}.
INLA exploits the conditional independence properties of the latent Gaussian field in order to dramatically reduce run time when approximating the posterior marginals.
An example of a scalable VB method is the work of \cite{tan2018gaussian}, where the Gaussian approximation is parameterised by a precision matrix instead of the standard covariance matrix; this cuts down on the number of variational parameters to be optimised in a mixed-effects regression setting.

The second challenge is the ability to distribute computation.
For large datasets, sometimes the only way to get inferences in a reasonable amount of time is to distribute the Bayesian inference across multiple computational nodes, and allowing the nodes to run in parallel.
Ideally this is able to be done recursively if more speed is required, leading to a divide-and-conquer approach to Bayesian computation.
Alternatively, the data may already be stored at different locations, and it is only feasible to perform Bayesian inference separately at each node before recombining results.
This could be because the data is too large to be contained in one machine, or because it comes from many different sources, all of which need to keep their data private \citep{li2020federated}.
In the Bayesian literature, there exist distributed versions of MCMC methods (for example, \citealt{ahn2014distributed}), VB methods (for example, \citealt{tran2016parallel}), and EP (for example, \citealt{hasenclever2017distributed}).
Among these approaches, only EP is currently able to distribute computation in a numerically stable way \citep{vehtari2020expectation}.
In particular, it is able to use the full prior for inference at each node (a fractional prior might not provide enough regularisation, leading to numerical instabilities---this undermines one of the key motivations of a Bayesian approach), while avoiding multiply counting the prior, which can lead to numerical instabilities when dividing by the prior at the end.

While both challenges have been separately resolved, there is currently no inference methodology for Bayesian mixed-effects regression models which is both scalable with respect to the number of groups and admits numerically stable distribution schemes, to the best of our knowledge.
We believe that addressing both issues are important when fitting such models to large datasets, and this motivated our development of such an approach based on the EP framework---this is the main contribution of this paper.

Why use EP as the Bayesian inference framework?
The primary reason is that we get numerically stable distribution for free, but there are other advantages to using EP for Bayesian inference.
To start, EP is typically much faster than MCMC methods \citep{minka2013expectation}.
In addition, it has a straightforward benefit over VB methods in that it does not require the posterior distribution to be differentiable; this has been exploited in works such as \cite{seeger2007bayesian} for lasso-based variable selection, and \cite{ko2016expectation} for regressions involving rectified linear functions.
One may also choose EP over INLA as it provides parametric approximations to the posterior, which can more easily be used in downstream analysis.
Furthermore, the latter method is known to be less accurate for binomial regression problems where the number of trials is small \citep{fong2010bayesian}.
EP, unlike the other approximate methods mentioned earlier, also provides an estimate of the predictive performance of its own approximation as a convenient byproduct of the algorithm \citep{qi2004predictive}, which can obviate the need for potentially expensive evaluation procedures such as cross-validation.

Despite these advantages, there are still some issues to solve if we choose to use the EP framework here.
Perhaps the main problem is that EP is not currently scalable with respect to the number of groups in a Bayesian mixed-effects regression setting.
While EP has been implemented for such a scenario \citep{kim2018expectation}, the largest variational parameter is $\mathcal{O}(L^2)$ in size, where $L$ is the number of groups, resulting in a computational complexity of $\mathcal{O}(L^2)$ for each pass.
Furthermore, \cite{cseke2011approximate} describe a sparse EP implementation for latent Gaussian models which in principle scales linearly with the number of groups, but their algorithm unfortunately requires multiple runs of EP over a grid of model hyperparameter values in a similar spirit to INLA; computation therefore becomes prohibitively slow for moderate to large amounts of hyperparameters.
We care about hyperparameters in this setting because many hyperparameters are needed to specify models with complex random-effects structures.
The other issue involves the derivation of the EP algorithm when there are multiple correlated random effects for each group.
In particular, refinements to the EP approximations of multivariate random-effects posterior factors currently seem to be intractable \citep{chen2020factor}; this precludes the usage of EP for a large subset of model types.

In this paper, we address the issues of scalability and tractability when implementing EP in a Bayesian mixed-effects regression setting (thus allowing for scalable and stably distributed Bayesian computation) in the specific case where only a single grouping factor is present in the data.
We focused on this simple case for brevity, but we expect that this approach generalises to more complicated random-effects structures.
The remainder of this paper is organised as follows.
In Section \ref{sec:ep}, EP in its standard implementation is briefly described.
In Section \ref{sec:mixed}, we outline the notation for the Bayesian mixed-effects regression model.
In Section \ref{sec:implement}, we describe our proposed EP implementation for such a model.
In Section \ref{sec:experiments}, numerical experiments are conducted to demonstrate the speed and accuracy of our proposed approach.
Finally, in Section \ref{sec:discussion}, we conclude with some discussion and suggestions for future research.

\subsection*{Notation}

For convenience, the general mathematical notation used throughout this paper is described here.
The notation $\mathcal{S}_{p}^+$ denotes the set of all $p\times p$ positive definite matrices.
The function $\text{dg}(\cdot)$ extracts the diagonal entries from its square matrix argument to form a vector, while the function $\text{diag}(\cdot)$ constructs the square matrix with diagonal entries equal to those of its vector argument.
Parentheses containing a list of comma-separated vectors denotes concatenation.
The functions $\phi_p(\cdot)$ and $\bar{\phi}_p(\cdot)$ are the densities of the $p$-dimensional multivariate Gaussian distribution, parameterised in terms of the mean and natural parameters respectively.
The function $w_p^{-1}(\cdot)$ is the density of the inverse-Wishart distribution on $p\times p$ matrices, parameterised in terms of the scale matrix and the degrees of freedom.
The functions $\text{rows}(\cdot)$ and $\text{cols}(\cdot)$ combine the rows and columns of their arguments respectively.
The notation $[\mA]_\vx$ for a covariance or precision matrix $\mA$ and an index vector $\vx$ indicates a submatrix selection of $\mA$ which corresponds to the elements of $\vx$.
For example, if $\mA$ is a $3\times3$ covariance matrix of the random vector $(X_1,X_2,X_3)$, then $\mA_{(X_1,X_3)}$ is the $2\times2$ matrix which contains as its entries (from left to right, top to bottom) $A_{11}$, $A_{13}$, $A_{31}$, and $A_{33}$.
A similar logic holds for the notation $[\vy]_\vx$, for a vector $\vy$ and an index vector $\vx$.

\section{EP basics}\label{sec:ep}

We briefly describe EP, and introduce key notation and nomenclature associated with the method.
EP is a message passing algorithm which acts on factor graphs, and was first introduced by \cite{minka2001family} as a way to perform ABI.
EP is closely related to VB methods \citep{blei2017variational}; both approaches aim to minimise a measure of divergence between the posterior distribution and its approximation.
The main difference between EP and VB is the structure of the approximation; while VB approaches iteratively refine a global approximation of the posterior approximation, EP iteratively refines local approximations to the factors of the posterior distribution.
The EP literature is extensive, and contains many excellent papers and tutorials describing the method; in addition to \cite{minka2001family}, examples include \cite{heskes2005approximate}, \cite{seeger2005expectation}, \cite{raymond2014expectation}, and \cite{vehtari2020expectation}.
In this section, we primarily describe the implementation of EP and put less emphasis on the theory.

Consider the approximate Bayesian inference problem of finding an approximation to a posterior distribution of the form
\begin{align}\label{eq:jlgen}
    p([\vtheta_j]_{j=1}^J|\sD)\propto\prod_{k=1}^Kf_k([\vtheta]_{j\in\mathcal{J}_k}),
\end{align}
where $\sD$ represents the data, $J$ is the number of parameters, $K$ is the number of factors (in the EP nomenclature these are known as \textit{sites}), $\vtheta_j\in\mathbb{R}^{d_j}$ are the model parameters (with $d_j$ being their dimensions), $\mathcal{J}_k=\{j\in[J]:f_k\text{ depends on }\vtheta_j\}$ is the index set of the parameters contained by $f_k$, $p:[\mathbb{R}^{d_j}]_{j=1}^J,\sD\to\mathbb{R}$ is the posterior distribution, and $f_k:[\mathbb{R}^{d_j}]_{j\in\mathcal{J}_k}\to\mathbb{R}$ are the sites with their dependencies on $\sD$ implicit.
EP approximates each site $f_k([\vtheta]_{j\in\mathcal{J}_k})$ by a product of (marginal) site approximations $\prod_{j\in\mathcal{J}_k}\widetilde{f}_{j,k}(\vtheta_j)$, where $\widetilde{f}_{j,k}$ is contained in some exponential family $\mathcal{Q}_j$.
This leads to a global approximation for the $\vtheta_j$ posterior marginal of the form $q_j(\vtheta_j)\propto\prod_{k\in\mathcal{K}_j}\widetilde{f}_{j,k}(\vtheta_j)$, where $\mathcal{K}_j=\{k\in[K]:f_k\text{ depends on }\vtheta_j\}$ is the index set of the sites which contain $\vtheta_j$.
The global EP approximation for the posterior distribution is then given by $q([\vtheta_j]_{j=1}^J)=\prod_{j=1}^Jq_j(\vtheta_j)$.
This approximation assumes independence between the $\vtheta_j$, so it is recommended to have as few parameters as possible; this can be achieved by concatenating parameters, for example.
The choice of the $\mathcal{Q}_j$ depends on the problem at hand, but typically a multivariate Gaussian family is used for unconstrained $\vtheta_j$ and an appropriately constrained family is used for constrained $\vtheta_j$.

In order to optimise the EP approximation, the site approximations are iteratively refined until convergence via Algorithm \ref{alg:epgen}.
Although convergence is not theoretically guaranteed \citep{hasenclever2017distributed}, it is often observed in practice.
We refer to each run through the for loop as a \textit{pass} through the site approximations.
The $\text{KL-proj}$ operator in line \ref{algline:klproj} denotes Kullback-Leibler (KL) projection, such that for a family of distributions $\mathcal{Q}$ and a distribution $p$, $\text{KL-proj}_{\mathcal{Q}}(p)\coloneqq\argmin_{q\in\mathcal{Q}}D_\text{KL}(p||q)$, with $D_\text{KL}(p||q)$ being the KL divergence from $p$ to $q$.
KL projection onto an exponential family is equivalent to moment matching the expectations of the sufficient statistics of that family.
The functions $f_{\backslash k}$ and $h_{j,k}$ are often referred to as the cavity and tilted distributions respectively in the EP nomenclature.
Note that it is also possible to update the global EP approximation (potentially multiple times) before a full pass has been made through the data.
While this approach is more computationally intensive, it also can reduce the amount of iterations until convergence \citep{barthelme2018divide}.

\begin{algorithm}
\caption{EP for approximating (\ref{eq:jlgen}).}\label{alg:epgen}
\begin{algorithmic}[1]
\Require Tractable exponential families $[\mathcal{Q}_j]_{j=1}^J$, sites $[f_k([\vtheta]_{j\in\mathcal{J}_k})]_{k=1}^K$, and site approximations $[[\widetilde{f}_{j,k}(\vtheta_j)]_{j\in\mathcal{J}_k}]_{k=1}^K$.
\State $q([\vtheta_j]_{j=1}^J)\propto\prod_{k=1}^K\prod_{j\in\mathcal{J}_k}\widetilde{f}_{j,k}(\vtheta_j)$ \label{algline:init}
\While {the $\widetilde{f}_{j,k}(\vtheta)$ have not converged}
\For {$(k',j')\in\{(k,j):k\in[K]\text{ and }j\in\mathcal{J}_k\}$} \label{algline:kj}
\State $f_{\backslash k'}([\vtheta_j]_{j=1}^J)\propto q([\vtheta_j]_{j=1}^J)/\prod_{j\in\mathcal{J}_{k'}}\widetilde{f}_{j,k'}(\vtheta_j)$ \label{algline:cavity}
\State $h_{j',k'}(\vtheta_{j'})\propto\int f_{k'}([\vtheta]_{j\in\mathcal{J}_{k'}})f_{\backslash k'}([\vtheta_j]_{j=1}^J)\,d\vtheta_{\backslash j'}$ \label{algline:tilted}
\State $\widetilde{f}_{j',k'}(\vtheta_{j'})\propto\text{KL-proj}_{\mathcal{Q}_{j'}}[h_{j',k'}](\vtheta_{j'})/\int f_{\backslash k'}([\vtheta_j]_{j=1}^J)\,d\vtheta_{\backslash j'}$ \label{algline:klproj}
\EndFor
\State $q([\vtheta_j]_{j=1}^J)\propto\prod_{k=1}^K\prod_{j\in\mathcal{J}_k}\widetilde{f}_{j,k}(\vtheta_j)$ \label{algline:update}
\EndWhile \\
\Return $q([\vtheta_j]_{j=1}^J)$
\end{algorithmic}
\end{algorithm}

As the bulk of the computation only requires the global approximation and local information, EP may be straightforwardly adapted for distributed data, where the sites are stored at different computational nodes.
In this setting, the computational nodes store the approximations associated with the sites they contain, and are responsible for their refinements (that is, the execution of the necessary for loop iterations in Algorithm \ref{alg:epgen}).
In addition, a central node is required to both run the remainder of the algorithm and store the global EP approximation.
Before each pass, the central node sends the updated global EP approximation to each computational node, and after the refinement of a site approximation, the updated version is sent back from its computational node to the central node; these are the only networking operations required in this framework.
More implementation details can be found in \cite{vehtari2020expectation}.
Note that the inference at each computational node effectively always involves the complete prior distribution, since the tilted distribution is an approximation of the complete (marginal) posterior distribution.
This allows for a full regularisation effect at each computational node, without the downside of having to divide by the prior distribution at the end of the algorithm.
Finally, it is clear that a computational node is able to recursively distribute its data and workload to additional sub-nodes if needed, further decreasing memory usage and run time.

\section{Bayesian mixed-effects regression}\label{sec:mixed}

The main focus in this paper is the Bayesian mixed-effects regression model with a single grouping factor; we outline the notation for such a model in this section.
The data consists of $N\in\mathbb{N}_{>0}$ observations partitioned into $L\in\mathbb{N}_{>0}$ groups, such that each observation consists of a response $y_n\in\mathbb{R}$, fixed-effects covariates $\vx_n\in\mathbb{R}^P$ (with $P\in\mathbb{N}_{\geq0}$), random-effects covariates $\vz_n\in\mathbb{R}^Q$ (with $Q\in\mathbb{N}_{>0}$), and additional data $\widetilde{\sD}_n$.
Let $\sD_n=\{y_n,\vx_n,\vz_n,\widetilde{\sD}_n\}$ be the data corresponding to the $n$th observation, $\sD=[\sD_n]_{n=1}^N$ be the overall data, $\vbeta\in\mathbb{R}^P$ be the vector of fixed effects, $\vu_l\in\mathbb{R}^Q$ be the vector of random effects for the $l$th group, $\vu=(\vu_1,\ldots,\vu_L)\in\mathbb{R}^{LQ}$ be the overall vector of random effects, $\mSigma\in\mathcal{S}_Q^+$ be the within-group covariance matrix of the random effects, and $\vgamma\in\mathbb{R}^{H}$ (with $H\in\mathbb{N}_{\geq0}$) be the model hyperparameters.
Let $\vtheta=(\vu,\vgamma,\vbeta)\in\mathbb{R}^{LQ+H+P}$ be the vector of all unconstrained parameters.
While it is possible to consider an unconstrained parameterisation of $\mSigma$ in terms of its Cholesky factor, the refinements of the random-effects sites for EP become difficult in this case, as this involves working with matrices where the entries of the Cholesky factorisation are randomly distributed.
The Bayesian mixed-effects model can be specified generally as
\begin{align*}
    y_n&\overset{\text{ind.}}{\sim} F(\vz_n^\tp\vu_{l(n)}+\vx_n^\tp\vbeta,\vgamma,\widetilde{\sD}_n)\ \text{for}\ n\in[N], \\
    \vu_l&\sim\mathcal{N}_Q(\vzero_Q,\mSigma)\ \text{for}\ l\in[L],\ \text{and} \\
    \vgamma,\vbeta,\mSigma&\sim H,
\end{align*}
where $l:[N]\to[L]$ returns the group number of the $n$th observation, and $F$ and $H$ are known distributions which depend on the specific model.
This corresponds to a posterior distribution of the form
\begin{align}\label{eq:mixed}
    p(\vtheta,\mSigma|\sD)\propto\left[\prod_{n=1}^Nf_n(\vtheta)\right]\left[\prod_{l=1}^Ls_l(\vtheta,\mSigma)\right]h(\vtheta,\mSigma),
\end{align}
where $f_n(\vtheta)=f^*(\vz_n^\tp\vu_{l(n)}+\vx_n^\tp\vbeta,\vgamma,y_n,\widetilde{\sD}_n)$, $s_l(\vtheta,\mSigma)=\abs{\mSigma}^{-1/2}\exp\left(-\vu_l^\tp\mSigma^{-1}\vu_l/2\right)$, and $h(\vtheta,\mSigma)$ is some joint prior density of $\vgamma$, $\vbeta$, and $\mSigma$.
Both $f^*$ and $h(\vtheta,\mSigma)$ are known functions and depend on the specific model.

\section{Scalable and tractable EP}\label{sec:implement}

In this section, we describe our EP implementation for approximating the Bayesian mixed-effects regression model defined by (\ref{eq:mixed}), modified in such a way so as to scale linearly with the number of groups $L$ and result in tractable refinements of the random-effects sites when $Q>1$.
A rough outline of this implementation, highlighting the main innovations, is as follows.
The site approximations are set to a combination of multivariate Gaussian and inverse-Wishart distributions, where the multivariate Gaussian site approximations are parameterised to only depend on the parameters which are contained in the corresponding site; this results in improved scalability of tilted inference at these site approximations.
We recognise that the precision matrix of the global multivariate Gaussian approximation admits structured sparsity; straightforward linear algebra may then be used to improve the scalability of updates and downdates involving this global approximation.
An adaptive version of power EP is used to simplify tilted inference at the Gaussian site approximations corresponding to the random-effects sites, and a moment propagation (MP) step is used to allow for tractable tilted inference at the inverse-Wishart site approximations.

We now give a more detailed description of the implementation.
To start, the global approximation has the form $q(\vtheta,\mSigma)=q_1(\vtheta)q_2(\mSigma)$ with
\begin{align}\label{eq:ep_approx}
    q_1(\vtheta)\propto\left[\prod_{n=1}^N\widetilde{f}_n(\vtheta)\right]\left[\prod_{l=1}^L\widetilde{s}_{l,1}(\vtheta)\right]\widetilde{h}_1(\vtheta)\quad\text{and}\quad q_2(\mSigma)\propto\left[\prod_{l=1}^L\widetilde{s}_{l,2}(\mSigma)\right]\widetilde{h}_2(\mSigma),
\end{align}
such that $\widetilde{f}_n(\vtheta)$, $\widetilde{s}_{l,1}(\vtheta)\widetilde{s}_{l,2}(\mSigma)$, and $\widetilde{h}_1(\vtheta)\widetilde{h}_2(\mSigma)$ approximate $f_n(\vtheta)$, $s_l(\vtheta,\mSigma)$, and $h(\vtheta,\mSigma)$ respectively for $n\in[N]$ and $l\in[L]$.
We use a multivariate Gaussian form for the site approximations in $\vtheta$, and an inverse-Wishart form for the site approximations in $\mSigma$, setting
\begin{align*}
    \widetilde{f}_n(\vtheta)&=\bar{\phi}_{Q+H+P}(\valpha_n;\mA_n\vr_{f,n}^*,\mA_{n}\mQ_{f,n}^*\mA_n^\tp), \\
    \widetilde{s}_{l,1}(\vtheta)&=\bar{\phi}_Q(\vu_l;\vr_{s_1,l},\mQ_{s_1,l}),\ \widetilde{s}_{l,2}(\mSigma)=w^{-1}_Q(\mSigma;\mPsi_{s_2,l},\nu_{s_2,l}), \\
    \widetilde{h}_1(\vtheta)&=\bar{\phi}_{H+P}(\vvarepsilon;\vr_{h_1},\mQ_{h_1}),\ \text{and}\ \widetilde{h}_2(\mSigma)=w^{-1}_Q(\mSigma;\mPsi_{h_2},\nu_{h_2}),
\end{align*}
where $\valpha_n=(\vu_{l(n)},\vgamma,\vbeta)\in\mathbb{R}^{Q+H+P}$ and $\vvarepsilon=(\vgamma,\vbeta)\in\mathbb{R}^{H+P}$.
Unlike a more standard EP implementation, we parameterise the site approximations in $\vtheta$ to only depend on the components of $\vtheta$ which are contained in its corresponding site; this reduces the time complexity of their refinement from quadratic to constant in the number of groups $L$, and is critical in achieving linear scaling.
Furthermore, a dimension reduction technique has been used for the Gaussian likelihood site approximations $\widetilde{f}_n(\vtheta)$ to reduce memory usage.
This technique takes advantage of the fact that the likelihood sites $f_n(\vtheta)$ partially depend on linear combinations of the model parameters; these linear combinations can be treated as single parameters in the context of Algorithm \ref{alg:epgen}; refer to \cite{chen2020factor}, \cite{vehtari2020expectation}, and \cite{zhou2023fast} for more details.
The quantities $\vr_{f,n}^*\in\mathbb{R}^{1+H}$ and $\mQ_{f,n}^*\in\mathcal{S}_{1+H}^+$ are the low-dimensional precision mean and precision matrix for the $n$th likelihood site approximation, and
\begin{align*}
    \mA_n=
    \begin{bmatrix}
        \widetilde{\vz}_n & \vzero_{Q\times H} \\
        \vzero_{H} & \mI_{H} \\
        \vx_n & \mI_{P\times H}
\end{bmatrix}\in\mathbb{R}^{(Q+H+P)\times(1+H)}
\end{align*}
is the corresponding matrix which converts between the low-dimensional and regular distributions.
It may be possible to perform additional dimension reduction for the model hyperparameters to further reduce run time.
Next, $\vr_{s_1,l}\in\mathbb{R}^Q$ and $\mQ_{s_1,l}\in\mathcal{S}_Q^+$ are the precision mean and precision matrix for the $l$th random effect site's $\vtheta$ component approximation, and $\mPsi_{s_2,l}\in\mathcal{S}_{Q}^+$ and $\nu_{s_2,l}>Q-1$ are the scale matrix and degrees of freedom for the $l$th random effect site's $\mSigma$ component approximation.
Finally, $\vr_{h_1}\in\mathbb{R}^{H+P}$ and $\mQ_{h_1}\in\mathcal{S}_{H+P}^+$ are the precision mean and precision matrix for the the joint prior site's $\vtheta$ component approximation, and $\mPsi_{h_2}\in\mathcal{S}_{Q}^+$ and $\nu_{h_2}>Q-1$ are the scale matrix and degrees of freedom for the joint prior site's $\mSigma$ component approximation.
In the following subsections, we describe how to implement the various components of the EP algorithm using this approximation structure, before outlining a distributed computation approach and performing a time complexity analysis.
We do not focus on the refinements of the joint prior site approximations, as the priors are usually chosen to have standard forms, and do not require an approximation in the first place.

\subsection{Initialisation}

This corresponds to line \ref{algline:init} of Algorithm \ref{alg:epgen}.
The global approximation $q_1(\vtheta)$ for the $\vtheta$ posterior marginal is seen to have a multivariate Gaussian form, whose parameters can be obtained from the $N+L+1$ site approximations in $\vtheta$.
Let $\mQ_\bullet\in\mathcal{S}_{LQ+H+P}^+$ and $\vr_\bullet\in\mathbb{R}^{LQ+H+P}$ be the precision matrix and precision mean respectively of $q_1(\vtheta)$.
In addition to these natural parameters, the mean parameters are also needed to compute the cavity distribution in line \ref{algline:cavity} of Algorithm \ref{alg:epgen}.
Let $\mSigma_\bullet\in\mathcal{S}_{LQ+H+P}^+$ and $\vmu_\bullet\in\mathbb{R}^{LQ+H+P}$ be the covariance matrix and mean respectively of $q_1(\vtheta)$.
On the other hand, the global approximation $q_2(\mSigma)$ for the $\mSigma$ posterior marginal is seen to have an inverse-Wishart form, whose parameters can be obtained from the $L+1$ site approximations in $\mSigma$ by using the fact that multiple inverse-Wishart densities may be combined as
\begin{align*}
    \prod_{i=1}^Iw_Q^{-1}(\mSigma;\mPsi_i,\nu_i)\propto w_Q^{-1}\left(\mSigma;\sum_{i=1}^I\mPsi_i,\sum_{i=1}^I\nu_i+(I-1)(Q+1)\right).
\end{align*}
Let $\mPsi_\bullet\in\mathcal{S}_{Q}^+$ and $\nu_\bullet>Q-1$ be the scale matrix and degrees of freedom respectively of $q_2(\mSigma)$; these represent the natural parameters.

Our main innovation in addressing scalability is recognising that $\mQ_\bullet$ admits structured sparsity; in particular, we have that
\begin{align*}
    \mQ_\bullet=
    \begin{bmatrix}
        \mB_{11} & \mB_{12} \\
        \mB_{12}^\tp & \mB_{22}
    \end{bmatrix},
\end{align*}
where $\mB_{11}=\text{bdiag}(\mB_{11,1},\ldots,\mB_{11,L})$.
In terms of dimensions, $\mB_{11}\in\mathbb{R}^{LQ\times LQ}$, $\mB_{11,l}\in\mathbb{R}^{Q\times Q}$ for $l\in[L]$, $\mB_{12}\in\mathbb{R}^{LQ\times(H+P)}$, and $\mB_{22}\in\mathbb{R}^{(H+P)\times(H+P)}$.
We aim to use this sparse representation of $\mQ_\bullet$ throughout the algorithm.
For convenience, decompose $\vr_\bullet$ in a similar way, with $\vr_\bullet=(\vd_1,\vd_2)$ where $\vd_1\in\mathbb{R}^{LQ}$ and $\vd_2=\mathbb{R}^{H+P}$.
It will be helpful to also define the auxiliary statistics $\widetilde{\mB}_{12}=\mB_{11}^{-1}\mB_{12}$, $\widetilde{\vd}_1=\widetilde{\mB}_{12}^\tp\vd_1$, $\bar{\mB}_{12}=\mB_{12}^\tp\widetilde{\mB}_{12}$, $\mS=\mB_{22}-\bar{\mB}_{12}$, and $\mT=\mS^{-1}$.
We have that $\mB_{11}^{-1}\mB_{12}=\text{rows}(\mB_{11,1}^{-1}\mB_{12,1},\ldots,\mB_{11,L}^{-1}\mB_{12,L})$.
In terms of dimensions, $\widetilde{\mB}_{12}\in\mathbb{R}^{LQ\times(H+P)}$, $\widetilde{\vd}_1\in\mathbb{R}^{H+P}$, and $\bar{\mB}_{12},\mS,\mT\in\mathbb{R}^{(H+P)\times(H+P)}$.

Using the new sparse representation of $\mQ_\bullet$ in addition to the auxiliary statistics, the covariance matrix $\mSigma_\bullet$ and mean $\vmu_\bullet$ of the global approximation for the $\vtheta$ posterior marginal may be written as
\begin{align}\label{eq:mean_par}
    \mSigma_\bullet=\mQ_\bullet^{-1}=
        \begin{bmatrix}
            \mB_{11}^{-1}+\widetilde{\mB}_{12}\mT\widetilde{\mB}_{12}^\tp & -\widetilde{\mB}_{12}\mT \\
            -\mT\widetilde{\mB}_{12}^\tp & \mT
        \end{bmatrix}\ \text{and}\ 
    \vmu_\bullet=\mQ_\bullet^{-1}\vr_\bullet=
        \begin{bmatrix}
            \mB_{11}^{-1}\vd_1-\widetilde{\mB}_{12}\mT(\vd_2-\widetilde{\vd}_1) \\
            \mT(\vd_2-\widetilde{\vd}_1)
        \end{bmatrix},
\end{align}
where $\mB_{11}^{-1}=\text{bdiag}(\mB_{11,1}^{-1},\ldots,\mB_{11,L}^{-1})$ and $\mB_{11}^{-1}\vd_1=(\mB_{11,1}^{-1}\vd_{1,1},\ldots,\mB_{11,L}^{-1}\vd_{1,L})$.
We aim to only calculate the parts of $\mSigma_\bullet$ and $\vmu_\bullet$ required for the task at hand throughout the algorithm.

\subsection{Refining likelihood site approximations}\label{sec:refine_like}

Such a refinement occurs when the $k'$ in line \ref{algline:kj} of Algorithm \ref{alg:epgen} corresponds to one of the $\widetilde{f}_n(\vtheta)$, for some $n\in[N]$.
First, we are required to perform a downdate of the global approximation in order to compute the cavity distribution $f_{\backslash n}(\vtheta)$; this corresponds to line \ref{algline:cavity}.
In this case, we actually require the low-dimensional cavity distribution as we are performing dimension reduction for the likelihood site approximations.
The downdate equations are the same as those in Algorithm 1 of \cite{zhou2023fast} (see Appendix A in the reference for derivations) with the exception that only the sections of $\mSigma_\bullet$ and $\vmu_\bullet$ corresponding to $\valpha_n$ are considered; these equations are
\begin{align}\label{eq:like_downdate}
    \begin{split}
        \mSigma_{\bullet,\valpha_n}\gets[\mSigma_\bullet]_{\valpha_n}\quad&\text{and}\quad\vmu_{\bullet,\valpha_n}\gets[\vmu_\bullet]_{\valpha_n}, \\
        \mSigma_{\bullet,n}^*\gets\mA_n^\tp\mSigma_{\bullet,\valpha_n}\mA_n\quad&\text{and}\quad\vmu_{\bullet,n}^*\gets\mA_n^\tp\vmu_{\bullet,\valpha_n}, \\
        \mQ_{\bullet,n}^*\gets(\mSigma_{\bullet,n}^*)^{-1}\quad&\text{and}\quad \vr_{\bullet,n}^*\gets \mQ_{\bullet,n}^*\vmu_{\bullet,n}^*, \\
        \mQ_{\backslash n}^*\gets \mQ_{\bullet,n}^*-\mQ_{f,n}^*\quad&\text{and}\quad \vr_{\backslash n}^*\gets \vr_{\bullet,n}^*-\vr_{f,n}^*,\ \text{and} \\
        \mSigma_{\backslash n}^*\gets(\mQ_{\backslash n}^*)^{-1}\quad&\text{and}\quad\vmu_{\backslash n}^*\gets\mSigma_{\backslash n}^*\vr_{\backslash n}^*.
    \end{split}
\end{align}
The quantities $\mSigma_{\bullet,\valpha_n}$ and $\vmu_{\bullet,\valpha_n}$ are the relevant parts of the mean parameters of the global approximation, $\mSigma_{\bullet,n}^*$ and $\vmu_{\bullet,n}^*$ are the mean parameters of the low-dimensional global approximation, $\mQ_{\bullet,n}^*$ and $\vr_{\bullet,n}^*$ are the corresponding natural parameters, $\mQ_{\backslash n}^*$ and $\vr_{\backslash n}^*$ are the natural parameters of the low-dimensional cavity distribution, and finally $\mSigma_{\backslash n}^*$ and $\vmu_{\backslash n}^*$ are the corresponding mean parameters.
We can show that
\begin{align*}
    \mSigma_{\bullet,\valpha_n}&=
    \begin{bmatrix}
        \mB_{11,l(n)}^{-1}+\widetilde{\mB}_{12,l(n)}\mT\widetilde{\mB}_{12,l(n)}^\tp & -\widetilde{\mB}_{12,l(n)}\mT \\
        -\mT\widetilde{\mB}_{12,l(n)}^\tp & \mT
    \end{bmatrix}\quad\text{and} \\
    \vmu_{\bullet,\valpha_n}&=
    \begin{bmatrix}
        \mB_{11,l(n)}^{-1}\vd_{1,l(n)}-\widetilde{\mB}_{12,l(n)}\mT(\vd_2-\widetilde{\vd}_1) \\
        \mT(\vd_2-\widetilde{\vd}_1)
    \end{bmatrix}.
\end{align*}

Following the downdate is the tilted inference step, which involves the computation of the tilted distribution (corresponding to line \ref{algline:tilted}) and the moments required for its KL projection onto the tractable exponential family (corresponding to line \ref{algline:klproj}).
As we are using a multivariate Gaussian family, we need to compute the mean and covariance of the tilted distribution in order to perform the KL projection.
Under the dimension reduction framework, these moments need to be computed for the low-dimensional tilted distribution
\begin{align*}
    h_{f,n}^*(\valpha^*)\propto f^*(\alpha_1^*,\valpha_{-1}^*,y_n,\widetilde{\sD}_n)\phi_{1+H}(\valpha^*;\vmu_{\backslash n}^*,\mSigma_{\backslash n}^*),
\end{align*}
where $\valpha^*\in\mathbb{R}^{1+H}$ is the low-dimensional parameter.
Let $\vmu_{h_{f,n}^*}\in\mathbb{R}^{1+H}$ and $\mSigma_{h_{f,n}^*}\in\mathcal{S}_{1+H}^+$ be the respective mean and covariance to be computed of the low-dimensional tilted distribution, and let $\widetilde{h}_{f,n}^*(\valpha^*)$ be its kernel.
Define
\begin{align*}
    I_{h_{f,n}^*}^0=\int\widetilde{h}_{f,n}^*(\valpha^*)\,d\valpha^*,\ \vI_{h_{f,n}^*}^1=\int\valpha^*\widetilde{h}_{f,n}^*(\valpha^*)\,d\valpha^*,\ \text{and}\ \mI_{h_{f,n}^*}^2=\int\valpha^*{\valpha^*}^\tp\widetilde{h}_{f,n}^*(\valpha^*)\,d\valpha^*
\end{align*}
to be the unnormalised zeroth, first, and second-order moments of the low-dimensional tilted distribution; these may be computed using $(1+H)$-dimensional numerical quadrature.
Ideally, $H$ is small such that the model does not have too many hyperparameters and the quadrature can be performed quickly.
Refer to \cite{vehtari2020expectation} for alternatives when $H$ is large.
Note that $H$ is not inclusive of the hyperparameters associated with the random effects, unlike for INLA or the EP implementation of \cite{cseke2011approximate} for example.
For some models (for example, heteroscedastic regression) it may be possible to perform additional dimension reduction for the model hyperparameters, such that only $D$-dimensional numerical quadrature, where $D<1+H$, is required to compute the unnormalised tilted distribution moments.
Once $I_{h_{f,n}^*}^0$, $\vI_{h_{f,n}^*}^1$, and $\mI_{h_{f,n}^*}^2$ have been calculated, we may recover the tilted distribution moments via
\begin{align*}
    \vmu_{h_{f,n}^*}=\frac{\vI_{h_{f,n}^*}^1}{I_{h_{f,n}^*}^0}\quad\text{and}\quad\mSigma_{h_{f,n}^*}=\frac{\mI_{h_{f,n}^*}^2}{I_{h_{f,n}^*}^0}-\vmu_{h_{f,n}^*}\vmu_{h_{f,n}^*}^\tp.
\end{align*}

The tilted inference step is succeeded by the update of the site approximation (corresponding to line \ref{algline:klproj}) and the update of the global approximation (corresponding to line \ref{algline:update}); when performing dimension reduction, the update equations are the same as those in Algorithm 1 of \cite{zhou2023fast} (again, see Appendix A in the reference for derivations), with the exception that we only update the sections of $\mQ_\bullet$ and $\vr_\bullet$ corresponding to $\valpha_n$; these equations are
\begin{align}\label{eq:like_update}
    \begin{split}
        \mQ_{h_{f,n}^*}\gets(\mSigma_{h_{f,n}^*})^{-1}\quad&\text{and}\quad \vr_{h_{f,n}^*}\gets \mQ_{h_{f,n}^*}\vmu_{h_{f,n}^*}, \\
        \widetilde{\mQ}_{f,n}^*\gets \mQ_{h_{f,n}^*}-\mQ_{\backslash n}^*\quad&\text{and}\quad\widetilde{\vr}_{f,n}^*\gets \vr_{h_{f,n}^*}-\vr_{\backslash n}^*, \\
        \Delta[\mQ_\bullet]_{\valpha_n}\gets\mA_n(\widetilde{\mQ}_{f,n}^*-\mQ_{f,n}^*)\mA_n^\tp\quad&\text{and}\quad\Delta[\vr_\bullet]_{\valpha_n}\gets\mA_n(\widetilde{\vr}_{f,n}^*-\vr_{f,n}^*), \\
        [\mQ_\bullet]_{\valpha_n}\gets[\mQ_\bullet]_{\valpha_n}+\Delta[\mQ_\bullet]_{\valpha_n}\quad&\text{and}\quad[\vr_\bullet]_{\valpha_n}\gets[\vr_\bullet]_{\valpha_n}+\Delta[\vr_\bullet]_{\valpha_n},\ \text{and} \\
        \mQ_{f,n}^*\gets \widetilde{\mQ}_{f,n}^*\quad&\text{and}\quad\widetilde{\vr}_{f,n}^*\gets\vr_{f,n}^*.
    \end{split}
\end{align}
The quantities $\mQ_{h_{f,n}^*}$ and $\vr_{h_{f,n}^*}$ are the natural parameters corresponding to the KL projection of the low-dimensional tilted distribution, $\widetilde{\mQ}_{f,n}^*$ and $\widetilde{\vr}_{f,n}^*$ are the updated natural parameters of the site approximation, and finally $\Delta[\mQ_\bullet]_{\valpha_n}$ and $\Delta[\vr_\bullet]_{\valpha_n}$ are the changes to the relevant parts of the natural parameters of the global approximation as a result of the site approximation update.
The second-last line of (\ref{eq:like_update}) may be written as
\begin{align*}
    \mB_{11,l(n)}&\gets\mB_{11,l(n)}+(\Delta[\mQ_\bullet]_{\valpha_n})_{1:Q,1:Q}, \\
    \mB_{12,l(n)}&\gets\mB_{12,l(n)}+(\Delta[\mQ_\bullet]_{\valpha_n})_{1:Q,-(1:Q)}, \\
    \mB_{22}&\gets\mB_{22}+(\Delta[\mQ_\bullet]_{\valpha_n})_{-(1:Q),-(1:Q)}, \\
    \vd_{1,l(n)}&\gets\vd_{1,l(n)}+(\Delta[\vr_\bullet]_{\valpha_n})_{1:Q},\ \text{and} \\
    \vd_2&\gets\vd_2+(\Delta[\vr_\bullet]_{\valpha_n})_{-(1:Q)}.
\end{align*}

\subsection{Refining random-effects site approximations}

Such a refinement occurs when the $k'$ and $j'$ in line \ref{algline:kj} of Algorithm \ref{alg:epgen} corresponds to one of the $\widetilde{s}_{l,1}(\vtheta)$ or $\widetilde{s}_{l,2}(\mSigma)$, for some $l\in[L]$.
Currently, the tilted inference step in these refinements is deemed to be intractable \citep{chen2020factor}.
However, we show that the random-effects site approximations may still be approximately updated in a principled and tractable way.

\subsubsection{Unconstrained site approximations}

First, consider the case where we are refining $\widetilde{s}_{l,1}(\vtheta)$, a random-effects site approximation in $\vtheta$, for some $l\in[L]$.
We propose that a power EP \citep{minka2004power} step is used to increase tractability in this scenario.
The power EP framework introduces a constant $\eta$, scales (by taking powers) the term in $\vtheta_{j'}$ in the denominator of line \ref{algline:cavity} and the terms not in $\vtheta_{j'}$ in line \ref{algline:tilted} by $\eta$, and scales the updated site approximation in line \ref{algline:klproj} by $1/\eta$.
Intuitively, power EP puts different weights on the cavity distribution and the site to build the tilted distribution, which can potentially lead to simpler calculations.
We set $\eta=-2/(\nu_{\backslash l}+1)$, where $\nu_{\backslash l}$ is calculated in (\ref{eq:Sigma_downdate}), for reasons which will become apparent during the tilted inference step.
Now, we are initially required to perform a downdate of the global approximation in order to compute the cavity distribution $s_{\backslash l}(\vtheta,\mSigma)$; this corresponds to line \ref{algline:cavity}.
The power EP downdate equations for $q_1(\vtheta)/\widetilde{s}_{l,1}(\vtheta)$, the $\vtheta$ component of the cavity distribution, are given by
\begin{align}\label{eq:theta_downdate}
    \begin{split}
        \mSigma_{\bullet,\vu_l}\gets[\mSigma_\bullet]_{\vu_l}\quad&\text{and}\quad\vmu_{\bullet,\vu_l}\gets[\vmu_\bullet]_{\vu_l}, \\
        \mQ_{\bullet,\vu_l}\gets\mSigma_{\bullet,\vu_l}^{-1}\quad&\text{and}\quad\vr_{\bullet,\vu_l}\gets\mQ_{\bullet,\vu_l}\vmu_{\bullet,\vu_l}, \\
        \mQ_{\backslash l}\gets\mQ_{\bullet,\vu_l}+[2/(\nu_{\backslash l}+1)]\mQ_{s_1,l}\quad&\text{and}\quad\vr_{\backslash l}\gets\vr_{\bullet,\vu_l}+[2/(\nu_{\backslash l}+1)]\vr_{s_1,l},\ \text{and} \\
        \mSigma_{\backslash l}\gets(\mQ_{\backslash l})^{-1}\quad&\text{and}\quad\vmu_{\backslash l}\gets\mSigma_{\backslash l}\vr_{\backslash l}.
    \end{split}
\end{align}
The form of the downdate equations is similar to those in Section \ref{sec:refine_like}, except there is no dimension reduction step and a power EP coefficient is added to the site parameters.
The quantities $\mSigma_{\bullet,\vu_l}$ and $\vmu_{\bullet,\vu_l}$ are the relevant sections of the mean parameters of the global approximation $q_1(\vtheta)$ for the $\vtheta$ posterior marginal, $\mQ_{\bullet,\vu_l}$ and $\vr_{\bullet,\vu_l}$ are the corresponding natural parameters, $\mQ_{\backslash l}$ and $\vr_{\backslash l}$ are the natural parameters of the $\vtheta$ component of the cavity distribution, and finally $\mSigma_{\backslash l}$ and $\vmu_{\backslash l}$ are the corresponding mean parameters.
We can show that
\begin{align}\label{eq:bullet_u_n}
    \mSigma_{\bullet,\vu_l}=\mB_{11,l}^{-1}+\widetilde{\mB}_{12,l}\mT\widetilde{\mB}_{12,l}^\tp\quad\text{and}\quad\vmu_{\bullet,\vu_l}=\mB_{11,l}^{-1}\vd_{1,l}-\widetilde{\mB}_{12,l}\mT(\vd_2-\widetilde{\vd}_1).
\end{align}
On the other hand, the power EP downdate equations for $q_2(\mSigma)/\widetilde{s}_{l,2}(\mSigma)$, the $\mSigma$ component of the cavity distribution, are given by
\begin{align}\label{eq:Sigma_downdate}
    \mPsi_{\backslash l}\gets\mPsi_\bullet-\mPsi_{s_2,l}\quad\text{and}\quad\nu_{\backslash l}\gets\nu_\bullet-\nu_{s_2,l}-(Q+1),
\end{align}
where $\mPsi_{\backslash l}$ and $\nu_{\backslash l}$ are the natural parameters of the $\mSigma$ component of the cavity distribution.

Next is the tilted inference step, which involves the computation of the tilted distribution (corresponding to line \ref{algline:tilted}) and the moments required for its KL projection onto the tractable exponential family (corresponding to line \ref{algline:klproj}).
The mean and covariance of the tilted distribution are required to perform the KL projection onto the multivariate Gaussian family.
Here, the power EP tilted distribution has the form
\begin{align*}
    h_{s_1,l}(\vu_l)\propto\left[\int s_l(\vtheta,\mSigma)w^{-1}_Q(\mSigma;\mPsi_{\backslash l},\nu_{\backslash l})\,d\mSigma\right]^{-2/(\nu_{\backslash l}+1)}\phi_Q(\vu_l;\vmu_{\backslash l},\mSigma_{\backslash l}).
\end{align*}
Let $\vmu_{h_{s_1,l}}\in\mathbb{R}^Q$ and $\mSigma_{h_{s_1,l}}\in\mathcal{S}_{Q}^+$ be its respective mean and covariance to be computed, and let $\widetilde{h}_{s_1,l}(\vu_l)$ be its kernel.
Define
\begin{align*}
    I_{h_{s_1,l}}^0=\int\widetilde{h}_{s_1,l}(\vu_l)\,d\vu_l,\ \vI_{h_{s_1,l}}^1=\int\vu_l\widetilde{h}_{s_1,l}(\vu_l)\,d\vu_l,\ \text{and}\ \mI_{h_{s_1,l}}^2=\int\vu_l\vu_l^\tp\widetilde{h}_{s_1,l}(\vu_l)\,d\vu_l
\end{align*}
to be the unnormalised zeroth, first, and second-order moments of the tilted distribution.
We first note that
\begin{align}\label{eq:C_s1_l}
    \begin{split}
        s_l(\vu_l,\mSigma)w^{-1}_Q(\mSigma;\mPsi_{\backslash l},\nu_{\backslash l})&=\abs{\mSigma}^{-1/2}\exp\left(-\frac{1}{2}\vu_l^\tp\mSigma^{-1}\vu_l\right)w^{-1}_Q(\mSigma;\mPsi_{\backslash l},\nu_{\backslash l}) \\
        &=C_l\exp\left(-\frac{1}{2}\vu_l^\tp\mSigma^{-1}\vu_l\right)w^{-1}_Q(\mSigma;\mPsi_{\backslash l},\nu_{\backslash l}+1) \\
        &=C_l\left[(1+\vu_l^\tp\mPsi_{\backslash l}^{-1}\vu_l)^{-(\nu_{\backslash l}+1)/2}\right]w^{-1}_Q(\mSigma;\mPsi_{\backslash l}+\vu_l\vu_l^\tp,\nu_{\backslash l}+1),
    \end{split}
\end{align}
where $C_l=[2^{Q/2}\Gamma_Q((\nu_{\backslash l}+1)/2)]/[\abs*{\mPsi_{\backslash l}}^{1/2}\Gamma_Q(\nu_{\backslash l}/2)]$ is a constant derived from the normalising constant of the inverse-Wishart distribution, and cancels out in the evaluation of $\vmu_{h_{s_1},l}$ and $\mSigma_{h_{s_1},l}$.
The function $\Gamma_Q$ is the multivariate gamma function with dimension $Q$.
The last line in (\ref{eq:C_s1_l}) may be derived by noting that $\vu_l^\tp\mSigma^{-1}\vu_l=\Tr(\vu_l\vu_l^\tp\mSigma^{-1})$ and subsequently using the matrix determinant lemma when manipulating the normalising constants.
Using (\ref{eq:C_s1_l}), we see that
\begin{align*}
    I_{h_{s_1,l}}^0&=C_l^{-2/(\nu_{\backslash l}+1)}\int(1+\vu_l^\tp\mPsi_{\backslash l}^{-1}\vu_l)\phi_Q(\vu_l;\vmu_{\backslash l},\mSigma_{\backslash l})\,d\vu_l, \\
    \vI_{h_{s_1,l}}^1&=C_l^{-2/(\nu_{\backslash l}+1)}\int\vu_l(1+\vu_l^\tp\mPsi_{\backslash l}^{-1}\vu_l)\phi_Q(\vu_l;\vmu_{\backslash l},\mSigma_{\backslash l})\,d\vu_l,\ \text{and} \\
    \mI_{h_{s_1,l}}^2&=C_l^{-2/(\nu_{\backslash l}+1)}\int\vu_l\vu_l^\tp(1+\vu_l^\tp\mPsi_{\backslash l}^{-1}\vu_l)\phi_Q(\vu_l;\vmu_{\backslash l},\mSigma_{\backslash l})\,d\vu_l.
\end{align*}
In order to evaluate each element of the above integrals, we are required to compute the Gaussian moments of quadratic forms and products of quadratic forms.
This is because for $i,j\in[Q]$, both $(\vu_l)_i$ and $(\vu_l)_i(\vu_l)_j$ can be considered quadratic forms in $\vu_l$; note that for $i,j\in[Q]$,
\begin{align*}
    (\vu_l)_i=\vu_l^\tp\mM_i\vu_l+\vm_i^\tp\vu_l+c_i\quad\text{and}\quad(\vu_l)_i(\vu_l)_j=\vu_l^\tp\mM_{i,j}\vu_l+\vm_{i,j}^\tp\vu_l+c_{i,j},
\end{align*}
where $\mM_i=\mzero_{Q\times Q}$, $\vm_i=\ve^Q_i$, $c_i=0$, $\mM_{i,j}=\frac{1}{2}(\mE^Q_{ij}+\mE^Q_{ji})$, $\vm_{i,j}=\vzero_{Q}$, and $c_{i,j}=0$, as required.
The cumulants of quadratic forms in Gaussian variables and the joint cumulants of products of quadratic forms in Gaussian variables are given in Theorem 3.3.2 and Theorem 3.3.3 respectively in \cite{mathai1992quadratic}.
Furthermore, recursions for the moments of a distribution based on its cumulants and for the joint moments of a multivariate distribution based on its joint cumulants are given in (5) and (10) respectively in \cite{smith1995recursive}.
Combining these results, we can show that
\begin{align*}
    I_{h_{s_1,l}}^0&=C_l^{-2/(\nu_{\backslash l}+1)}(1+\Tr(\mPsi_{\backslash l}^{-1}\mSigma_{\backslash l})+\vmu_{\backslash l}^\tp\mPsi_{\backslash l}^{-1}\vmu_{\backslash l}), \\
    \vI_{h_{s_1,l}}^1&=C_l^{-2/(\nu_{\backslash l}+1)}\left[\left(1+\Tr(\mPsi_{\backslash l}^{-1}\mSigma_{\backslash l})+\vmu_{\backslash l}^\tp\mPsi_{\backslash l}^{-1}\vmu_{\backslash l}\right)\vmu_{\backslash l}+2\mSigma_{\backslash l}\mPsi_{\backslash l}^{-1}\vmu_{\backslash l }\right],\ \text{and} \\
    \mI_{h_{s_1,l}}^2&=C_l^{-2/(\nu_{\backslash l}+1)}\left[\left(1+\Tr(\mPsi_{\backslash l}^{-1}\mSigma_{\backslash l})+\vmu_{\backslash l}^\tp\mPsi_{\backslash l}^{-1}\vmu_{\backslash l}\right)\left(\mSigma_{\backslash l}+\vmu_{\backslash l}\vmu_{\backslash l}^\tp\right)\right. \\
    &\qquad\qquad\qquad\ \ \left.+2\left(\mSigma_{\backslash l}\mPsi_{\backslash l}^{-1}\mSigma_{\backslash l}+\mSigma_{\backslash l}\mPsi_{\backslash l}^{-1}\vmu_{\backslash l}\vmu_{\backslash l}^\tp+\vmu_{\backslash l}\vmu_{\backslash l}^\tp\mPsi_{\backslash l}^{-1}\mSigma_{\backslash l}\right)\right].
\end{align*}
We may then recover the tilted distribution moments via
\begin{align*}
    \vmu_{h_{s_1,l}}=\frac{\vI_{h_{s_1,l}}^1}{I_{h_{s_1,l}}^0}
    \quad\text{and}\quad
    \mSigma_{h_{s_1,l}}=\frac{\mI_{h_{s_1,l}}^2}{I_{h_{s_1,l}}^0}-\vmu_{h_{s_1,l}}\vmu_{h_{s_1,l}}^\tp.
\end{align*}

Finally, we have the update of the site approximation (corresponding to line \ref{algline:klproj}) and the update of the global approximation (corresponding to line \ref{algline:update}); in this case, the power EP update equations are given by
\begin{align}\label{eq:theta_update}
    \begin{split}
        \mQ_{h_{s_1,l}}\gets(\mSigma_{h_{s_1,l}})^{-1}\quad&\text{and}\quad\vr_{h_{s_1,l}}\gets\mQ_{h_{s_1,l}}\vmu_{h_{s_1,l}}, \\
        \widetilde{\mQ}_{s_1,l}\gets-[(\nu_{\backslash l}+1)/2](\mQ_{h_{s_1,l}}-\mQ_{\backslash l})\quad&\text{and}\quad\widetilde{\vr}_{s_1,l}\gets-[(\nu_{\backslash l}+1)/2](\vr_{h_{s_1,l}}-\vr_{\backslash l}), \\
        \Delta[\mQ_\bullet]_{\vu_l}\gets\widetilde{\mQ}_{s_1,l}-\mQ_{s_1,l}\quad&\text{and}\quad\Delta[\vr_\bullet]_{\vu_l}\gets\widetilde{\vr}_{s_1,l}-\vr_{s_1,l}, \\
        [\mQ_\bullet]_{\vu_l}\gets[\mQ_\bullet]_{\vu_l}+\Delta[\mQ_\bullet]_{\vu_l}\quad&\text{and}\quad[\vr_\bullet]_{\vu_l}\gets[\vr_\bullet]_{\vu_l}+\Delta[\vr_\bullet]_{\vu_l},\ \text{and} \\
        \mQ_{s_1,l}\gets\widetilde{\mQ}_{s_1,l}\quad&\text{and}\quad\vr_{s_1,l}\gets\widetilde{\vr}_{s_1,l}.
    \end{split}
\end{align}
The form of the update equations is similar to those in Section \ref{sec:refine_like}, except there is no dimension reduction component and a power EP coefficient is added to the site parameters.
The quantities $\mQ_{h_{s_1,l}}$ and $\vr_{h_{s_1,l}}$ are the natural parameters corresponding to the KL projection of the tilted distribution, $\widetilde{\mQ}_{s_1,l}$ and $\widetilde{\vr}_{s_1,l}$ are the updated natural parameters of the site approximation, and finally $\Delta[\mQ_\bullet]_{\vu_l}$ and $\Delta[\vr_\bullet]_{\vu_l}$ are the changes to the relevant parts of the natural parameters of the global approximation as a result of the site approximation update.
The second-last line of (\ref{eq:theta_update}) may be written as
\begin{align*}
    \mB_{11,l}\gets\mB_{11,l}+\Delta[\mQ_\bullet]_{\vu_l}\quad\text{and}\quad\vd_{1,l}&\gets\vd_{1,l}+\Delta[\vr_\bullet]_{\vu_l}.
\end{align*}

\subsubsection{Constrained site approximations}

Now consider the case where we are refining $\widetilde{s}_{l,2}(\mSigma)$, a random-effects site approximation in $\mSigma$, for some $l\in[L]$.
We propose an MP \citep{ormerod2022moment} step to update all of $\widetilde{s}_{l,2}(\mSigma)$ simultaneously, under the assumption that the joint prior distribution can be written as $h(\vtheta,\mSigma)=h^*(\vtheta)w_Q^{-1}(\mSigma;\mPsi_\mSigma,\nu_\mSigma)$, where $\mPsi_\mSigma\in\mathcal{S}_{Q}^+$ and $\nu_\mSigma>Q-1$.
In other words, the $\mSigma$ component factors out of the joint prior distribution and has an inverse-Wishart form; we believe that this is a relatively mild assumption.
Consider the full conditional posterior distribution with respect to $\mSigma$,
\begin{align*}
    p(\mSigma|\vtheta,\sD)\propto\left[\prod_{l=1}^Ls_l(\vtheta,\mSigma)\right]w_Q^{-1}(\mSigma;\mPsi_\mSigma,\nu_\mSigma)=w_Q^{-1}\left(\mSigma;\mPsi_\mSigma+\sum_{l=1}^L\vu_l\vu_l^\tp,\nu_\mSigma+L\right).
\end{align*}
Under an MP update scheme, matching statistics are calculated for this full conditional distribution, expectations are taken with respect to $q_1(\vtheta)$, the current approximation of the conditioning parameters, and these are then matched with the corresponding quantities of a standard distribution to get the updated version of $q_2(\mSigma)$.
An inverse-Wishart form was chosen so as to match the refinements of the random-effects site approximations in $\vtheta$, where it is assumed that the corresponding approximations in $\mSigma$ have inverse-Wishart forms.
To simplify the matching procedure, the matching statistics were chosen to be $\mOmega=\mathbb{E}_{p(\mSigma)}(\mSigma)$ and $\omega=\sum_{i=1}^Q\mathbb{V}_{p(\mSigma)}(\mSigma_{ii})$, where $p(\mSigma)$ is either the full conditional posterior distribution or the inverse-Wishart distribution.
In the former case, write the matching statistics as $\mOmega(\vtheta)$ and $\omega(\vtheta)$ to make their dependence on $\vtheta$ explicit.
It can be shown that
\begin{align*}
    \mOmega(\vtheta)=\frac{\mPsi_\mSigma+\sum_{l=1}^L\vu_l\vu_l^\tp}{\nu_\mSigma+L-Q-1}\quad\text{and}\quad\omega(\vtheta)=\frac{2\sum_{i=1}^Q\left[(\mPsi_\mSigma)_{i,i}+\sum_{l=1}^L(\vu_l)_i^2\right]^2}{(\nu_\mSigma+L-Q-1)^2(\nu_\mSigma+L-Q-3)}.
\end{align*}
We are now required to take expectations with respect to $q_1(\vtheta)$.
For the quantity $\mOmega(\vtheta)$, this is straightforward, with
\begin{align*}
    \mathbb{E}_{q_1(\vtheta)}\left[\mOmega(\vtheta)\right]=\frac{\mPsi_\mSigma+\sum_{l=1}^L([\mSigma_\bullet]_{\vu_l}+[\vmu_\bullet]_{\vu_l}[\vmu_\bullet]_{\vu_l}^\tp)}{\nu_\mSigma+L-Q-1}.
\end{align*}
For the quantity $\omega(\vtheta)$, we perform the expansion
\begin{align*}
    \left((\mPsi_\mSigma)_{i,i}+\sum_{l=1}^L(\vu_l)_i^2\right)^2=(\mPsi_\mSigma)_{i,i}^2+2(\mPsi_\mSigma)_{i,i}\left(\sum_{l=1}^L(\vu_l)_i^2\right)+\left(\sum_{l=1}^L(\vu_l)_i^2\right)^2
\end{align*}
and note that $\sum_{l=1}^L(\vu_l)_i^2$ is a quadratic form in Gaussian random variables.
More precisely, we have that $\sum_{l=1}^L(\vu_l)_i^2=\vx^\tp\mA\vx$, where $$\vx\sim\mathcal{N}_L((([\vmu_\bullet]_{\vu_1})_i,\ldots,([\vmu_\bullet]_{\vu_L})_i),\text{diag}(([\mSigma_\bullet]_{\vu_1})_{i,i},\ldots,([\mSigma_\bullet]_{\vu_L})_{i,i}))$$ and $\mA=\mI_L$.
Using the formulas for the first and second moments and some algebra, we can then show that $\mathbb{E}_{q_1(\vtheta)}\left[\omega(\vtheta)\right]=$
\begin{align*}
    \frac{2\sum_{i=1}^Q\left[\left(\sum_{l=1}^L2([\mSigma_\bullet]_{\vu_l})_{i,i}^2+4([\mSigma_\bullet]_{\vu_l})_{i,i}([\vmu_\bullet]_{\vu_l})_i^2\right)+\left((\mPsi_\mSigma)_{i,i}+\sum_{l=1}^L([\mSigma_\bullet]_{\vu_l})_{i,i}+([\vmu_\bullet]_{\vu_l})_i^2\right)^2\right]}{(\nu_\mSigma+L-Q-1)^2(\nu_\mSigma+L-Q-3)}.
\end{align*}
Equation (\ref{eq:bullet_u_n}) should be used to evaluate $[\mSigma_\bullet]_{\vu_l}$ and $[\vmu_\bullet]_{\vu_l}$ in the preceding formulas.
Once the expectations of the matching statistics have been computed, it is easy to see that the inverse-Wishart distribution with $\mOmega=\mathbb{E}_{q_1(\vtheta)}\left[\mOmega(\vtheta)\right]$ and $\omega=\mathbb{E}_{q_1(\vtheta)}\left[\omega(\vtheta)\right]$ (and thus the updated version of $q_2(\mSigma)$) has parameters
\begin{align*}
    \mPsi_\bullet=\left(\frac{2\sum_{i=1}^Q(\mathbb{E}_{q_1(\vtheta)}\left[\mOmega(\vtheta)\right])_{i,i}^2}{\mathbb{E}_{q_1(\vtheta)}\left[\omega(\vtheta)\right]}+2\right)\mathbb{E}_{q_1(\vtheta)}\left[\mOmega(\vtheta)\right]\quad\text{and}\quad\nu_\bullet=\frac{2\sum_{i=1}^Q(\mathbb{E}_{q_1(\vtheta)}\left[\mOmega(\vtheta)\right])_{i,i}^2}{\mathbb{E}_{q_1(\vtheta)}\left[\omega(\vtheta)\right]}+Q+3.
\end{align*}
Finally, the $\widetilde{s}_{l,2}(\mSigma)$ may be updated by splitting the new $\mPsi_\bullet$ and $\nu_\bullet$ equally among these approximations, after the contribution from the prior distribution has been accounted for.
More precisely,
\begin{align*}
    \mPsi_{s_2,l}\gets(\mPsi_\bullet-\mPsi_\mSigma)/L\quad\text{and}\quad\nu_{s_2,l}\gets(\nu_\bullet-\nu_\mSigma)/L - Q - 1
\end{align*}
for $l\in[L]$.
For simplicity, we recommend that the MP step be performed after the refinement of all other site approximations in $\vtheta$ and the update of the parameters of $q_1(\vtheta)$ in the current pass.

\subsection{Inference after convergence}

Once EP has converged, we may straightforwardly perform inference on $\mSigma$ using $q_2(\mSigma)$ (for example, we may calculate posterior expectations).
For inference on $\vtheta$, there are two main scalable options.
The first is to calculate the marginal distributions of $q_1(\vtheta)$; 
we may use (\ref{eq:mean_par}) to evaluate both $\text{dg}(\mSigma_\bullet)=([\text{dg}(\mB_{11,l}^{-1}+\widetilde{\mB}_{12,l}\mT\widetilde{\mB}_{12,l}^\tp)]_{l=1}^L,\text{dg(\mT)})$ and $\vmu_\bullet$, given the parameters of $q_1(\vtheta)$ as well as their auxiliary statistics.
We may also generate samples from $q_1(\vtheta)$ (these may be used to infer properties of $\vtheta$).
Start by decomposing $\mQ_\bullet$ into $\mL(\mQ_\bullet)\mL(\mQ_\bullet)^\tp$; we have
\begin{align*}
    \mL(\mQ_\bullet)=
    \begin{bmatrix}
        \mL(\mB_{11}) & \mzero_{LQ\times(H+P)} \\
        \mB_{12}^\tp\mL(\mB_{11})^{-\tp} & \mL(\mS)
    \end{bmatrix}\quad\text{and}\quad
    \mL(\mQ_\bullet)^{-1}=
    \begin{bmatrix}
        \mL(\mB_{11})^{-1} & \mzero_{LQ\times(H+P)} \\
        -\mL(\mS)^{-1}\widetilde{\mB}_{12}^\tp & \mL(\mS)^{-1}
    \end{bmatrix},
\end{align*}
where
\begin{align*}
    \mL(\mB_{11})&=\text{bdiag}\left(\mL(\mB_{11,1}),\ldots,\mL(\mB_{11,L})\right), \\
    \mB_{12}^\tp\mL(\mB_{11})^{-\tp}&=\text{cols}\left(\mB_{12,1}^\tp\mL(\mB_{11,1})^{-\tp},\ldots,\mB_{12,L}^\tp\mL(\mB_{11,L})^{-\tp}\right),\ \text{and} \\
    \mL(\mB_{11})^{-1}&=\text{bdiag}\left(\mL(\mB_{11,1})^{-1},\ldots,\mL(\mB_{11,L})^{-1}\right).
\end{align*}
Both the Cholesky factor and its inverse admit the same sparse structure as $\mQ_\bullet$.
A sample $\vtheta_\text{sam.}$ from $q_1(\vtheta)$ may then be obtained using
\begin{align*}
\vtheta_\text{sam.}=\mL(\mQ_\bullet)^{-\tp}(\mL(\mQ_\bullet)^{-1}\vr_\bullet)+\mL(\mQ_\bullet)^{-\tp}\vz,    
\end{align*}
where the entries of $\vz\in\mathbb{R}^{LQ+H+P}$ are independent standard univariate Gaussians.
Note that for a general vector $\vv=(\vv_1,\vv_2)$ with $\vv_1\in\mathbb{R}^{LQ}$ and $\vv_2\in\mathbb{R}^{H+P}$,
\begin{align*}
    \mL(\mQ_\bullet)^{-1}\vv=
    \begin{bmatrix}
        \mL(\mB_{11})^{-1}\vv_1 \\
        \mL(\mS)^{-1}(\vv_2-\widetilde{\mB}_{12}^\tp\vv_1)
    \end{bmatrix}\quad\text{and}\quad
    \mL(\mQ_\bullet)^{-\tp}\vv=
    \begin{bmatrix}
        \mL(\mB_{11})^{-\tp}\vv_1-\widetilde{\mB}_{12}\mL(\mS)^{-\tp}\vv_2 \\
        \mL(\mS)^{-\tp}\vv_2
    \end{bmatrix},
\end{align*}
where
\begin{align*}
    \mL(\mB_{11})^{-1}\vv_1&=\left(\mL(\mB_{11,1})^{-1}\vv_{1,1},\ldots,\mL(\mB_{11,L})^{-1}\vv_{1,L}\right)\ \text{and} \\
    \mL(\mB_{11})^{-\tp}\vv_1&=\left(\mL(\mB_{11,1})^{-\tp}\vv_{1,1},\ldots,\mL(\mB_{11,L})^{-\tp}\vv_{1,L}\right).
\end{align*}

\subsection{Distributed computation}

In Section \ref{sec:ep}, the process of adapting standard EP to a distributed setting for a general model was described.
We now give some additional points regarding distributed Bayesian computation when using our proposed EP implementation on mixed-effects regression models in particular.
Typically, it is only the likelihood sites that are split up across multiple computational nodes; therefore, the computational nodes will only contain the likelihood sites and their approximations, while the remaining sites/approximations are stored at the central node.
Of course, if the reason for the distribution was to decrease run time/memory usage, then it is also perfectly acceptable split up the random-effects and prior sites across computational nodes.
As for distributing the MP step, since it does not require any local information, all associated computations should be performed at the central node, although the updated site approximations will need to be sent out to the computational nodes.
Within this step, it is also possible to parallelise/distribute the summations required to compute the expectations with respect to $q_1(\vtheta)$, so as to further speed up the algorithm.
Finally, inference for $\vtheta$ and $\mSigma$ after EP has converged only requires global information, and therefore should be run on the central node.

\subsection{Time complexity}

We perform a basic time complexity analysis of our proposed EP implementation to show that it does indeed scale linearly with respect to the number of groups $L$.
Assume that $N$ and $L$ vary, with the remaining variables set to constants.
During the initialisation stage, the site approximations $\widetilde{f}_n(\vtheta)$, $\widetilde{s}_{l,1}(\vtheta)$, and $\widetilde{s}_{l,2}(\mSigma)$ are first initialised in $\mathcal{O}(N)$ time.
Next, the global approximation parameters $\mB_{11}$, $\mB_{12}$, $\mB_{22}$, $\vd_1$, $\mPsi_\bullet$, and $\nu_\bullet$ are initialised in $\mathcal{O}(N)$ time.
Finally, the auxiliary statistics $\widetilde{\mB}_{12}$, $\widetilde{\vd}_1$, $\bar{\mB}_{12}$, $\mS$, and $\mT$ are initialised in $\mathcal{O}(L)$ time.
Given the global approximation parameters and the auxiliary statistics, the refinement of the $\widetilde{f}_n(\vtheta)$ and $\widetilde{s}_{l,1}(\vtheta)$ each take $\mathcal{O}(1)$ time, so that a refinement of all the $\widetilde{f}_n(\vtheta)$ and $\widetilde{s}_{l,1}(\vtheta)$ takes $\mathcal{O}(N)$ time.
After these refinements, the parameters of $q_1(\vtheta)$ as well as their auxiliary statistics are be recomputed; this requires $\mathcal{O}(N+L)$ time, similar to the initialisation stage.
Given these quantities, the MP step then takes $\mathcal{O}(L)$ time.
Each subsequent pass then takes the same amount of time, such that if we assume that the proposed EP implementation converges after some fixed number of passes through all site approximations (which is typical), then the overall time complexity for convergence is $\mathcal{O}(N+L)$.
After convergence, it is clear that inference for both $\vtheta$ and $\mSigma$ can be performed in $\mathcal{O}(L)$ time; in the former case we are able to exploit structured sparsity in $\mQ_\bullet$ to either quickly calculate the marginals or perform a fast Cholesky decomposition.
Thus, the overall run time of our proposed EP implementation is $\mathcal{O}(N+L)$. 
In most scenarios, each group contains a fixed number of observations such that $N\propto L$ approximately.
This reduces the time complexity to $\mathcal{O}(L)$, as required.

\section{Experiments}\label{sec:experiments}

Numerical experiments were conducted to demonstrate the speed and accuracy of our proposed EP implementation (call this EP-S, where S stands for sparse).
We examined two popular model types: zero-inflated Poisson regression and binomial regression.
Two sets of numerical experiments were conducted for each model type.
Firstly, the run time of EP-S was compared with a non-scalable version where the structured sparsity in $\mQ_\bullet$ is not accounted for in the calculations (call this EP-NS; this is comparable to the approach of \citealt{kim2018expectation}) on simulated datasets; we aimed to show that using the former approach results in a significant reduction in computational time in practice.
Secondly, both the accuracy and run time of EP-S were compared to those of INLA \citep{rue2009approximate} and Gaussian variational Bayes parameterised in terms of the precision matrix (\citealt{tan2018gaussian}---call this GVB) for smaller to medium-sized real-world datasets; these alternatives were chosen as they also scale linearly with respect to the number of groups.
Note however that INLA only approximates the posterior of the $\mSigma$ components and model hyperparameters over a discrete grid, whereas EP-S and GVB return continuous approximations.
Here, we aimed to show that EP-S has comparable accuracies and run times compared to current state-of-the-art approaches for scalable inference in Bayesian mixed-effects regression models.
For binomial regression, we additionally considered a larger real-world dataset where distributed Bayesian computation may be necessary, and compared both the approximation and run time of a distributed implementation of EP-S to that of INLA.
The code for these experiments can be found at \href{https://github.com/jackson-zhou-sydney/EP-MM}{https://github.com/jackson-zhou-sydney/EP-MM}.

Both EP-S and EP-NS were implemented in \CC; we damped the updates to the site approximation parameters by factors of 0.5 and 0.8 for the simulated and real-world datasets respectively, and set the minimum and maximum number of passes through the sites to be 5 and 100 respectively.
Damped updates have been seen in the EP literature (see \citealt{seeger2005expectation} and \citealt{vehtari2020expectation} for example), and are typically used to help the global approximations to remain proper.
Meanwhile, the minimum pass constraint prevents premature stopping which can lead to reduced accuracies, and the maximum pass constraint limits run time.
Within this range, convergence was determined (with the algorithm terminated) if for all types of site approximation parameters, the maximum change across site approximations in the current pass was less than 0.05 times the corresponding average of the four passes.
Change was measured by the absolute value, the Euclidean norm, and the Frobenius norm (for scalars, vectors, and matrices respectively) of the difference.
INLA was implemented via the R-INLA R package \citep{rue2009approximate}, which runs C code internally.
The default settings were used, and samples were generated from the joint posterior approximation using the \texttt{inla.posterior.sample} function.
GVB was implemented using the Julia code obtained from the authors of \cite{tan2018gaussian}.
We adopted the settings used there, and set the maximum number of iterations to 100,000 so as to limit run time.
It is of our opinion that the implementations of EP, INLA, and GVB (based on \CC, C, and Julia respectively) were generally able to be compared fairly in terms of run time.

Where possible, MCMC implemented via the \texttt{cmdstanr} R package \citep{cmdstanr} was used as the gold standard to evaluate accuracy throughout the experiments.
MCMC was run with ten independent chains, with each chain containing 1000 warm-up iterations and 10000 sampling iterations; the default values were chosen for the remaining settings.
Convergence was verified by checking that the $\widehat{R}$ statistic was less than 1.1 for all parameters, as per the recommendation in \cite{gelman1995bayesian}---this was always the case.
After convergence, for each marginal component $i$ we computed $\text{dev}(\widehat{\mu}_i)\coloneqq(\widehat{\mu}_i-\mu_i)/\sigma_i$, $\text{adev}(\widehat{\mu}_i)\coloneqq\abs{\text{dev}(\widehat{\mu}_i)}$, $\text{dev}(\widehat{\sigma}_i)\coloneqq\widehat{\sigma}_i/\sigma_i$, and $\text{adev}(\widehat{\sigma}_i)\coloneqq\text{dev}(\widehat{\sigma}_i)$ if $\text{dev}(\widehat{\sigma}_i)\geq1$ (else $1/\text{dev}(\widehat{\sigma}_i)$), where $\widehat{\mu}_i$ and $\widehat{\sigma}_i$ are the $i$th estimated marginal mean and SD respectively of the method to be evaluated, while $\mu_i$ and $\sigma_i$ are those of MCMC.
Accuracy was evaluated by averaging these deviations and absolute deviations across certain groups of marginal components (e.g., all marginal components, or only those associated with $\vbeta$); denote these averages by $\overline{\text{dev}(\widehat{\mu}_i)}$, $\overline{\text{adev}(\widehat{\mu}_i)}$, $\overline{\text{dev}(\widehat{\sigma}_i)}$, and $\overline{\text{adev}(\widehat{\sigma}_i)}$ respectively.
For the first two quantities, we use the arithmetic mean, while for the last two quantities, we use the geometric mean.
Higher marginal accuracy is indicated by values of $\overline{\text{adev}(\widehat{\mu}_i)}$ closer to zero and values of $\overline{\text{adev}(\widehat{\sigma}_i)}$ closer to one.
Furthermore, the values of $\overline{\text{dev}(\widehat{\mu}_i)}$ and $\overline{\text{dev}(\widehat{\sigma}_i)}$ allow us to diagnose over/underestimation of the posterior mean and/or variance.
Multivariate accuracy was also evaluated, but the corresponding results were omitted from the main text for conciseness as they mostly agreed with the marginal accuracies.
More details can be found in Appendix \ref{sec:more_results}.

On the other hand, the run time for a method was measured as the time in seconds needed to generate marginal approximations for all parameters, in addition to 1,000 samples from the joint approximation for the posterior distribution.
Five repetitions were performed, with the mean and standard deviation of the run time across these repetitions being measured.

\subsection{Zero-inflated Poisson regression}

First, consider a Bayesian mixed-effects zero-inflated Poisson (ZIP) regression model with a log link function, such that $y_n\in\mathbb{N}_{\geq0}$ and $\widetilde{\sD}_n=o_n$ for $n\in[N]$ (the offset for that observation), $\vgamma=\lambda$ (the logit of the probability of a structural zero), and the likelihood function $f_n(\vtheta)$ for the $n$th observation is given by
\begin{footnotesize}
\begin{align}\label{eq:zip}
    \begin{split}
        f_n(\vtheta)&=f^*(\vz_n^\tp\vu+\vx_n^\tp\vbeta,\lambda,y_n,o_n) \\
        &=\begin{cases}
            \text{expit}(\lambda)+[1-\text{expit}(\lambda)]\exp[-\exp(\vz_n^\tp\vu+\vx_n^\tp\vbeta+o_n)] & \text{if }y_n=0, \\
            [1-\text{expit}(\lambda)]\exp[y_n(\vz_n^\tp\vu+\vx_n^\tp\vbeta+o_n)-\exp(\vz_n^\tp\vu+\vx_n^\tp\vbeta+o_n)]/y_n! & \text{if }y_n>0.
        \end{cases}
    \end{split}
\end{align}
\end{footnotesize}
It is also possible to consider a model where the probability of a structural zero depends on certain covariates--- dimension reduction may be used to speed up the refinements in this case.
Assume that the joint prior distribution has the natural form
\begin{align*}
    h(\vtheta,\mSigma)&=\phi(\lambda;\mu_\lambda,\Sigma_\lambda)\phi_P(\vbeta;\vmu_\vbeta,\mSigma_\vbeta)w_Q^{-1}(\mSigma;\mPsi_\mSigma,\nu_\mSigma),
\end{align*}
where $\mu_\lambda\in\mathbb{R}$, $\Sigma_\lambda\in\mathbb{R}_{>0}$, $\vmu_\vbeta\in\mathbb{R}^{P}$, $\mSigma_\vbeta\in\mathcal{S}_{P}^+$, $\mPsi_\mSigma\in\mathcal{S}_{Q}^+$, and $\nu_\mSigma>Q-1$.

For both EP-S and EP-NS, the joint prior site approximations were initialised to be exact approximations of the joint prior sites, and did not need to be iteratively refined throughout the algorithm.
The parameters of the non--joint prior site approximations were initialised as $\vr_{f,n}^*=\vzero_2$, $\mQ_{f,n}^*=\mI_2$, $\vr_{s_1,l}=\vzero_Q$, $\mQ_{s_1,l}=\mI_Q$, $\mPsi_{s_2,l}=\mI_Q$, and $\nu_{s_2,l}=Q+2$.
Additional EP derivations for the ZIP model can be found in Appendix \ref{sec:model_ep}.

\begin{figure}
    \begin{center}
        \includegraphics{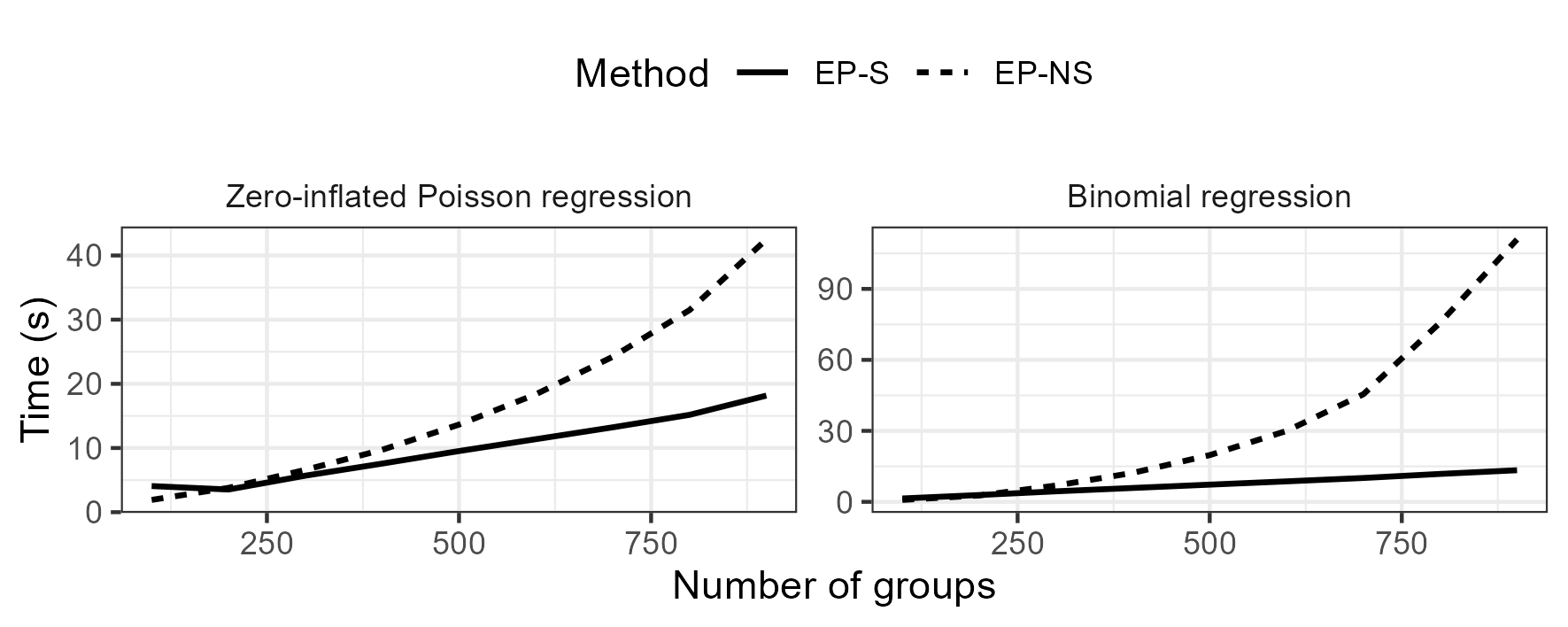}
    \end{center}
    \caption{Mean run times in seconds of EP-S and EP-NS when performing inference on a Bayesian mixed-effects zero-inflated Poisson/binomial regression model. Standard deviations in run times were always less than 5 seconds.}
    \label{fig:scalability_plot}
\end{figure}

In the first set of experiments, we considered simulated datasets with $L=100,200,\ldots,900$.
We set the number of observations per group at ten (so that $N=10L$), $P=8$, and $Q=2$.
The true values of the parameters were set as $\vbeta=0.25\cdot(1,-1,1,-1,1,-1,1,-1)$, $\lambda=\text{logit}(0.05)$, and $\mSigma=0.5\mI_Q$, with $\vu$ simulated according to (\ref{eq:mixed}).
In terms of the data, for all $n\in[N]$ the entries in $\vx_n$ and $\vz_n$ were simulated from i.i.d.\ $\mathcal{N}(0,1)$ distributions, with the exception of the first entries which were set to one to correspond to the intercept.
Furthermore, for all $n\in[N]$ we set $o_n=0$ and simulate $y_n$ according to (\ref{eq:zip}).
For the parameters of the prior distributions, we set $\mu_\kappa=0$, $\Sigma_\kappa=10000$, $\mu_\lambda=0$, $\Sigma_\lambda=10000$, $\vmu_\vbeta=\vzero_P$, $\mSigma_\vbeta=10000\mI_P$, $\mPsi_\mSigma=\mI_Q$, and $\nu_\mSigma=Q+2$.
The results for the first set of experiments are shown in Figure \ref{fig:scalability_plot}, where it can be seen that the run times of EP-S are much lower than those of EP-NS in practice.
Furthermore, the run times of EP-S and EP-NS roughly scale linearly and polynomially respectively, which agrees with the analysis from earlier in the paper.

In the second set of experiments, the datasets used were \texttt{Epilepsy} ($N=295$, $P=4$, $L=59$, $Q=1$), \texttt{Fly} ($N=1437$, $P=12$, $L=479$, $Q=1$), \texttt{Owl\_1} ($N=599$, $P=6$, $L=27$, $Q=1$), \texttt{Owl\_3} ($N=599$, $P=4$, $L=27$, $Q=3$), \texttt{Salamanders} ($N=644$, $P=4$, $L=23$, $Q=4$), and \texttt{Webworms} ($N=1300$, $P=2$, $L=13$, $Q=2$).
Dataset descriptions can be found in Appendix \ref{sec:data}.
The marginal absolute deviations and run times are shown in Table \ref{table:zip_abs_marginal}, with corresponding marginal deviations (non-absolute), multivariate accuracies, and marginal distribution plots given in Appendix \ref{sec:more_results}.
INLA tended to have the best absolute deviations in marginal means and SDs (i.e., absolute deviations in marginal means closest to zero and absolute deviations in marginal SDs closest to one), followed very closely by EP-S and then GVB.
All methods approximated the posterior marginals fairly well, with it generally being the case that $\overline{\text{adev}(\widehat{\mu}_i)}\leq0.2$ and $\overline{\text{adev}(\widehat{\sigma}_i)}\leq1.2$.
Across all methods, the approximations for the $\lambda$ and $\text{vech}(\mSigma)$ marginal components were generally worse compared to those of the $\vu$ and $\vbeta$ marginal components.
In terms of run time, EP-S was the fastest method, followed by INLA and then GVB.
Overall, EP-S had comparable accuracies and shorter run times compared to the other scalable ABI approaches.

\begin{table}
\centering
\fontsize{9pt}{9pt}\selectfont
\begin{tabular}{@{}cccccccc@{}}
\toprule
\multirow{2}{*}{Dataset}     & \multirow{2}{*}{Method} & \multicolumn{5}{c}{$\overline{\text{adev}(\widehat{\mu}_i)}$/$\overline{\text{adev}(\widehat{\sigma}_i)}$} & \multirow{2}{*}{Time (s)} \\
                             &                         & $(\vtheta,\text{vech}(\mSigma))$       & $\vu$       & $\lambda$      & $\vbeta$      & $\text{vech}(\mSigma)$      &                           \\ \midrule
\multirow{3}{*}{\texttt{Epilepsy}}         & EP-S                      & 0.04/1.03   & 0.04/1.02   & 0.44/1.34  & 0.03/1.03  & 0.00/1.08  & 0.2 $\pm$ 0.0             \\
                             & INLA                    & 0.06/1.02   & 0.06/1.02   & 0.18/1.07  & 0.04/1.01  & 0.07/1.06  & 4.8 $\pm$ 0.2             \\
                             & GVB                     & 0.05/1.02   & 0.05/1.02   & 0.10/1.02  & 0.09/1.01  & 0.14/1.10  & 11.5 $\pm$ 0.2             \\ \midrule
\multirow{3}{*}{\texttt{Fly}} & EP-S                      & 0.04/1.02   & 0.04/1.02   & 0.02/1.04  & 0.19/1.20  & 0.09/1.19  & 1.3 $\pm$ 0.0             \\
                             & INLA                    & 0.06/1.02   & 0.06/1.02   & 0.04/1.09  & 0.05/1.13  & 0.05/1.04  & 6.5 $\pm$ 0.3             \\
                             & GVB                     & 0.06/1.09   & 0.05/1.08   & 0.35/1.02  & 0.22/1.83  & 0.18/1.04  & 235.8 $\pm$ 2.7             \\ \midrule
\multirow{3}{*}{\texttt{Owl\_1}} & EP-S                      & 0.04/1.03   & 0.04/1.02   & 0.02/1.01  & 0.03/1.03  & 0.18/1.17  & 0.4 $\pm$ 0.0             \\
                             & INLA                    & 0.04/1.03   & 0.04/1.03   & 0.07/1.09  & 0.02/1.02  & 0.15/1.09  & 5.1 $\pm$ 0.3             \\
                             & GVB                     & 0.05/1.06   & 0.03/1.06   & 0.04/1.01
  & 0.10/1.04  & 0.29/1.29  & 21.8 $\pm$ 0.2             \\ \midrule
\multirow{3}{*}{\texttt{Owl\_3}}         & EP-S                      & 0.09/1.03 & 0.08/1.02 & 0.15/1.02 & 0.04/1.01 & 0.20/1.22 & 0.7 $\pm$ 0.0             \\
                             & INLA                    & 0.06/1.04 & 0.06/1.04 & 0.07/1.05 & 0.04/1.01 & 0.07/1.09 & 11.3 $\pm$ 0.20             \\
                             & GVB                     & 0.17/1.1 & 0.07/1.07 & 0.18/1.05 & 0.06/1.06 & 1.58/1.62 & 20.3 $\pm$ 0.1             \\ \midrule
\multirow{3}{*}{\texttt{Salamanders}} & EP-S                      & 0.13/1.08 & 0.11/1.05 & 0.06/1.00 & 0.16/1.10 & 0.27/1.45 & 1.1 $\pm$ 0.0             \\
                             & INLA                    & 0.09/1.08 & 0.08/1.05 & 0.01/1.01 & 0.10/1.06 & 0.17/1.33 & 14.9 $\pm$ 0.7             \\
                             & GVB                     & 0.22/1.38 & 0.12/1.27 & 0.01/1.01 & 0.11/1.36 & 1.14/3.00 & 27.6 $\pm$ 0.3             \\ \midrule
\multirow{3}{*}{\texttt{Webworms}}    & EP-S                      & 0.05/1.06 & 0.05/1.05 & 0.06/1.09 & 0.04/1.07 & 0.02/1.11 & 1.4 $\pm$ 0.0             \\
                             & INLA                    & 0.04/1.04 & 0.04/1.02 & 0.05/1.02 & 0.05/1.04 & 0.11/1.29 & 4.9 $\pm$ 0.0             \\
                             & GVB                     & 0.08/1.13 & 0.05/1.08 & 0.08/1.09 & 0.07/1.16 & 0.32/1.62 & 22.6 $\pm$ 0.2             \\ \bottomrule
\end{tabular}
\caption{Marginal absolute deviations from MCMC and run times (mean $\pm$ standard deviation in seconds) of ABI methods when performing inference on a Bayesian mixed-effects ZIP regression model, for smaller to medium-sized datasets.}
\label{table:zip_abs_marginal}
\end{table}

\subsection{Binomial regression}

Now consider a Bayesian mixed-effects binomial regression model with a probit link function, such that $y_n\in\mathbb{N}_{\geq0}$ and $\widetilde{\sD}_n=\aleph_n\in\mathbb{N}_{>0}$ (the number of binomial trials for that observation) for $n\in[N]$, there are no model hyperparameters so that $H=0$ and $\vgamma$ does not exist, and the likelihood function $f_n(\vtheta)$ for the $n$th observation is given by
\begin{align}\label{eq:binom}
    f_n(\vtheta)=f^*(\vz_n^\tp\vu+\vx_n^\tp\vbeta,y_n,\aleph_n)=\binom{\aleph_n}{y_n}\Phi(\vz_n^\tp\vu+\vx_n^\tp\vbeta)^{y_n}[1-\Phi(\vz_n^\tp\vu+\vx_n^\tp\vbeta)]^{\aleph_n-y_n}.
\end{align}
Assume that the joint prior distribution has the natural form
\begin{align*}
    h(\vtheta,\mSigma)=\phi_P(\vbeta;\vmu_\vbeta,\mSigma_\vbeta)w_Q^{-1}(\mSigma;\mPsi_\mSigma,\nu_\mSigma),
\end{align*}
where $\vmu_\vbeta\in\mathbb{R}^{P}$, $\mSigma_\vbeta\in\mathcal{S}_{P}^+$, $\mPsi_\mSigma\in\mathcal{S}_{Q}^+$, and $\nu_\mSigma>Q-1$.

As with the ZIP model, the joint prior site approximations for EP and its non-scalable versions were initialised to be exact approximations of the joint prior sites, and did not need to be iteratively refined throughout the algorithm.
The parameters of the non--joint prior site approximations were initialised as $r_{f,n}^*=0$, $Q_{f,n}^*=1$, $\vr_{s_1,l}=\vzero_Q$, $\mQ_{s_1,l}=\mI_Q$, $\mPsi_{s_2,l}=\mI_Q$, and $\nu_{s_2,l}=Q+2$.
Additional EP derivations for the binomial model can be found in Appendix \ref{sec:model_ep}.

In the first set of experiments, we considered simulated datasets with $L=100,200,\ldots,900$.
As with the ZIP model, we set the number of observations per group at ten (so that $N=10L$), $P=8$, and $Q=2$.
The true values of the parameters were set as $\vbeta=(1,-1,1,-1,1,-1,1,-1)$ and $\mSigma=0.5\mI_Q$, with $\vu$ simulated according to (\ref{eq:mixed}).
In terms of the data, for all $n\in[N]$ the entries in $\vx_n$ and $\vz_n$ were simulated from i.i.d.\ $\mathcal{N}(0,1)$ distributions, with the exception of the first entries which were set to one to correspond to the intercept.
Furthermore, for all $n\in[N]$ we set $\aleph_n=1$ and simulate $y_n$ according to (\ref{eq:binom}).
For the parameters of the prior distributions, we set $\vmu_\vbeta=\vzero_P$, $\mSigma_\vbeta=10000\mI_P$, $\mPsi_\mSigma=\mI_Q$, and $\nu_\mSigma=Q+2$.
The results for the first set of experiments are shown in Figure \ref{fig:scalability_plot}, where it can be seen that the run time of EP-S scales linearly, compared to the polynomial scaling of EP-NS.
Additionally, there is more of a difference in run times between the two methods for the binomial model, compared to the ZIP model.
This is because now a smaller proportion of the run time for both methods is spent performing tilted inference when refining the likelihood site approximations, as only univariate numerical quadrature is required compared to bivariate numerical quadrature for the ZIP model, due to the extra hyperparameter $\lambda$ in the ZIP model---see Appendix \ref{sec:model_ep}.

In the second set of experiments, the datasets used were \texttt{CTSIB} ($N=480$, $P=8$, $L=40$, $Q=1$), \texttt{HDP\_1} ($N=381$, $P=4$, $L=35$, $Q=1$), \texttt{HDP\_3} ($N=381$, $P=3$, $L=35$, $Q=3$), \texttt{Salamanders} ($N=644$, $P=4$, $L=23$, $Q=4$), \texttt{Tapped} ($N=715$, $P=3$, $L=23$, $Q=2$), and \texttt{Toenail} ($N=1908$, $P=4$, $L=294$, $Q=1$).
Dataset descriptions can be found in Appendix \ref{sec:data}.
The marginal absolute deviations and run times are shown in Table \ref{table:zip_abs_marginal}, with corresponding marginal deviations (non-absolute), multivariate accuracies, and marginal distribution plots given in Appendix \ref{sec:more_results}.
EP-S tended to have the best absolute deviations in marginal means and SDs (i.e., absolute deviations in marginal means closest to zero and absolute deviations in marginal SDs closest to one), followed by INLA and GVB, which tended to have similar absolute deviations.
Much like the ZIP model, all methods approximated the posterior marginals fairly well (though perhaps to less of an extent compared to the ZIP case), with it generally being the case that $\overline{\text{adev}(\widehat{\mu}_i)}\leq0.2$ and $\overline{\text{adev}(\widehat{\sigma}_i)}\leq1.2$.
Similar to before, the approximations for the $\text{vech}(\mSigma)$ marginal components were generally worse compared to those of the $\vu$ and $\vbeta$ marginal components.
It was particularly hard to approximate the posterior associated with the \texttt{Toenail} dataset, where the absolute deviations in both marginal means and SDs were moderately worse compared to the other datasets.
The subpar accuracy of INLA for \texttt{CTSIB}, \texttt{Tapped}, and \texttt{Toenail} may be explained by the small numbers of binomial trials corresponding to each data point in these datasets; INLA is known to underperform for binomial data corresponding to small numbers of binomial trials \citep{fong2010bayesian}.
In terms of run time, EP-S was the fastest method, followed by INLA and then GVB (similar to the ZIP case).
Overall, EP-S had comparable (if not better) accuracies and shorter run times compared to the other scalable ABI approaches.

\begin{table}[h!]
\centering
\fontsize{9pt}{9pt}\selectfont
\begin{tabular}{@{}cccccccc@{}}
\toprule
\multirow{2}{*}{Dataset}     & \multirow{2}{*}{Method} & \multicolumn{4}{c}{$\overline{\text{adev}(\widehat{\mu}_i)}$/$\overline{\text{adev}(\widehat{\sigma}_i)}$} & \multirow{2}{*}{Time (s)} \\
                             &                         & $(\vtheta,\text{vech}(\mSigma))$       & $\vu$ & $\vbeta$      & $\text{vech}(\mSigma)$      &                           \\ \midrule
\multirow{3}{*}{\texttt{CTSIB}}         & EP-S                      & 0.06/1.06 & 0.06/1.03 & 0.06/1.12 & 0.12/1.77 & 0.3 $\pm$ 0.2             \\
                             & INLA                    & 0.24/1.11 & 0.19/1.08 & 0.50/1.15 & 0.23/2.08 & 1.7 $\pm$ 0.1             \\
                             & GVB                     & 0.10/1.15 & 0.09/1.12 & 0.12/1.19 & 0.44/1.77 & 10.3 $\pm$ 0.4             \\ \midrule
\multirow{3}{*}{\texttt{HDP\_1}} & EP-S                      & 0.04/1.02 & 0.04/1.02 & 0.03/1.02 & 0.04/1.05 & 0.5 $\pm$ 0.2             \\
                             & INLA                    & 0.03/1.02 & 0.03/1.02 & 0.02/1.02 & 0.06/1.03 & 2.0 $\pm$ 0.0             \\
                             & GVB                     & 0.04/1.02 & 0.03/1.02 & 0.07/1.03 & 0.25/1.18 & 13.6 $\pm$ 0.2             \\ \midrule
\multirow{3}{*}{\texttt{HDP\_3}} & EP-S                      & 0.05/1.02 & 0.05/1.02 & 0.02/1.01 & 0.08/1.14 & 0.1 $\pm$ 0.0             \\
                             & INLA                    & 0.04/1.02 & 0.04/1.02 & 0.01/1.01 & 0.07/1.06 & 7.3 $\pm$ 0.0             \\
                             & GVB                     & 0.22/1.07 & 0.07/1.04 & 0.16/1.08 & 2.86/1.87 & 13.1 $\pm$ 0.1             \\ \midrule
\multirow{3}{*}{\texttt{Salamanders}}         & EP-S                      & 0.04/1.07 & 0.04/1.03 & 0.10/1.01 & 0.04/1.49 & 0.1 $\pm$ 0.0             \\
                             & INLA                    & 0.06/1.06 & 0.05/1.04 & 0.06/1.03 & 0.15/1.28 & 11.9 $\pm$ 0.4             \\
                             & GVB                     & 0.18/1.32 & 0.08/1.22 & 0.08/1.26 & 1.05/2.65 & 15.7 $\pm$ 0.2             \\ \midrule
\multirow{3}{*}{\texttt{Tapped}} & EP-S                      & 0.03/1.07 & 0.03/1.05 & 0.05/1.03 & 0.03/1.60 & 0.2 $\pm$ 0.0             \\
                             & INLA                    & 0.14/1.06 & 0.12/1.05 & 0.49/1.03 & 0.09/1.26 & 2.9 $\pm$ 0.0             \\
                             & GVB                     & 0.11/1.17 & 0.09/1.14 & 0.13/1.08 & 0.40/1.97 & 17.1 $\pm$ 0.1             \\ \midrule
\multirow{3}{*}{\texttt{Toenail}}    & EP-S                      & 0.12/1.14 & 0.12/1.13 & 0.19/1.14 & 0.89/2.74 & 1.1 $\pm$ 0.0             \\
                             & INLA                    & 0.55/1.23 & 0.54/1.23 & 1.12/1.17 & 1.46/1.65 & 2.4 $\pm$ 0.1             \\
                             & GVB                     & 0.08/1.11 & 0.08/1.11 & 0.11/1.17 & 0.75/2.16 & 75.5 $\pm$ 0.7             \\ \bottomrule
\end{tabular}
\caption{Marginal absolute deviations from MCMC and run times (mean $\pm$ standard deviation in seconds) of ABI methods when performing inference on a Bayesian mixed-effects binomial regression model, for smaller to medium-sized datasets.}
\label{table:binom_abs_marginal}
\end{table}

In addition to the previous experiments which focus on smaller to medium-sized real-world datasets, we also evaluated EP-S on a larger real-world dataset where distributed computation may be necessary.
We aimed to demonstrate that a distributed implementation of EP-S both returns similar approximations and runs in a comparable amount of time compared to (non-distributed) INLA.
In this setting, MCMC was excluded due to its large run time (around 1.15 minutes required per iteration) and GVB was excluded due to its moderately high run time (around 0.9 seconds required per iteration) and the need for it to be tuned and rerun for convergence.
Both marginal deviations and multivariate accuracy now used INLA instead of MCMC as the reference distribution.

The dataset of interest relates to the United States Bureau of Labor Statistics' National Longitudinal Survey of Youth (NLSY, \citealt{moore2000national}).
We focused on the 1997 cohort, with the data available at \href{https://www.nlsinfo.org/content/cohorts/nlsy97}{https://www.nlsinfo.org/content/cohorts/nlsy97}.
At the time of writing, the data consisted of yearly survey responses of 8,984 United States participants from the years 1997 through to 2022 inclusive.
The participants were aged from 12 through to 16 inclusive at the time of their recruitment in 1997, and were the grouping factor in the data (each participant has multiple observations, each corresponding to a single year).
Though the data was downloaded from a single location, it is entirely possible that it needs to be split up across multiple different locations for privacy/logistical reasons.
For example, participants living in different states could have their data stored on state data repositories which are not easily merged due to privacy legislation.
We manually split the participants into eight groups of roughly equal sizes, and ran distributed EP-S across these groups (based on the descriptions given in Section \ref{sec:ep} and Section \ref{sec:implement}).
A large range of questions are asked in the survey; example question categories include education, employment, geography, attitudes, and health.
We aimed to infer the effect of various covariates on a participant's health.
This was determined by their response to the question ``how is the respondent's general health?''.
The possible responses were excellent, very good, good, fair and poor, and the model aimed to predict if the response was either excellent or very good for a given combination of participant and year.
The specific covariates used for the analysis are described in Appendix \ref{sec:data}.
After filtering to exclude missing data, we have $N=25856$, $P=205$, $L=4269$, and $Q=3$.
EP-S was terminated after exactly 100 passes through the data to ensure that convergence was reached across the large amount of fixed effects compared to the previous examples.

The marginal absolute deviations and run times are shown in Table \ref{table:binom_abs_marginal_big}, with corresponding marginal deviations (non-absolute), multivariate accuracies, and marginal distribution plots given in Appendix \ref{sec:more_results}.
The run times of non-distributed EP-S and INLA were 1015.1 seconds and 3650.9 seconds respectively.
The EP-S approximation is seen to generally agree with the INLA approximation, with the exception of the $\text{vech}(\mSigma)$ components, where there is a two SD gap between the mean estimates and three-fold gap between the SD estimates.
Examination of the marginal deviations (non-absolute) in Appendix \ref{sec:more_results} reveals that EP-S was overestimating the marginal posterior mean and underestimating the marginal posterior SD when compared to INLA.
Given that INLA tended to underestimate the marginal posterior SD for the smaller datasets (this can again be seen in Appendix \ref{sec:more_results}), this indicates that EP may also be underestimating the marginal posterior SD in this case.
This can be explained by the independence assumption made between $\vtheta$ and $\mSigma$ in the construction of the approximating form; such a problem has been observed for mean-field variational approximations (see for example \citealt{ormerod2022moment}).
On the other hand, INLA tended to underestimate the marginal posterior mean for the smaller datasets; the degree of overestimation of the marginal posterior mean of EP-S is then most likely less than that indicated by the current results.
In terms of run time, distributed EP-S was 3.4 times faster compared to the non-distributed version, which itself was almost four times faster than INLA.
We expect the speed-up of distribution to be greater for larger $N$ or for models where the tilted inference at each site is more computationally demanding (such as the ZIP model).
In both these cases, the networking overhead associated with distributed computation comprises a smaller proportion of the total run time.

\begin{table}
\centering
\fontsize{10pt}{10pt}\selectfont
\begin{tabular}{@{}cccccccc@{}}
\toprule
\multicolumn{4}{c}{$\overline{\text{adev}(\widehat{\mu}_i)}$/$\overline{\text{adev}(\widehat{\sigma}_i)}$} & \multirow{2}{*}{Time (s)} \\
                            $(\vtheta,\text{vech}(\mSigma))$       & $\vu$ & $\vbeta$      & $\text{vech}(\mSigma)$      &                           \\ \midrule
                             0.16/1.23 & 0.16/1.23 & 0.24/1.06 & 2.04/3.52 & 301.1             \\ \bottomrule
\end{tabular}
\caption{Marginal absolute deviations from INLA and run time of distributed EP-S when performing inference on a Bayesian mixed-effects binomial regression model for the NLSY dataset.}
\label{table:binom_abs_marginal_big}
\end{table}

\section{Discussion} \label{sec:discussion}

In this paper, we developed a Bayesian inference framework for mixed-effects regression models (with a single grouping factor) that is both scalable with respect to the number of groups and is numerically stable when distributed.
EP was used as the underlying methodology, with the main technical innovations being the sparse representation of the global EP approximation and an MP-based refinement for multivariate random effect factor approximations.
We demonstrated our framework on ZIP regression and binomial regression examples (for both smaller and larger datasets) where it was shown to have comparable accuracies with INLA and GVB, while requiring moderately less run time.
There are several limitations of our scalable and distributed framework, however.
Firstly, accuracy in the approximation of the marginal posteriors of the $\mSigma$ components deteriorates for larger datasets, though this is a non-issue if only the fixed effects are of interest.
Secondly, our approach does not scale with the number of hyperparameters $H$; recall that $(1+H)$-dimensional numerical quadrature is required to refine a likelihood site approximation.
This may be dealt with by using dimension reduction (this depends on the specific model), or approximating the tilted inference step using say a Laplace approximation \citep{eskin2003laplace}.
Finally, EP has non-negligible overhead (unlike INLA and GVB) in terms of the derivations required to implement the algorithm.
While there exists literature on automating the tilted inference step \citep{jitkrittum2015kernel}, accuracy can often be compromised in these cases.
Natural extensions to our work in this paper include implementations for more complicated random-effects structures (such as crossed and nested random effects) and more general models which can use random effects (such as generalised additive models).

\section*{Acknowledgements}

The authors would like to thank Dr.\ Alistair Senior from the Sydney Precision Data Science Centre for his generous contribution of \texttt{Fly}, one of the datasets used for the numerical experiments.
In addition, the following sources of funding are gratefully acknowledged: the Australian Government Research Training Program Scholarship to the first author and the Australian Research Council Discovery Project Grant (DP210100521) to the second author.

\appendix
\renewcommand{\thesection}{\Alph{section}}

\section{Model-specific EP derivations}\label{sec:model_ep}

This section contains model-specific EP derivations for the models considered in Section \ref{sec:experiments}.
In particular, the form of the low-dimensional tilted distribution when performing dimension reduction, along with how inference may be performed, is described.

\subsection{Zero-inflated Poisson regression}

For a refinement of a likelihood site approximation $\widetilde{f}_n$, the kernel of the low-dimensional tilted distribution is given by
\begin{align*}
    \widetilde{h}_{f,n}^*(\valpha^*)&=f^*(\alpha_1^*,\alpha_2^*,y_n)\phi_2(\valpha^*;\vmu_{\backslash n}^*,\mSigma_{\backslash n}^*) \\
    &=\begin{cases}
        \{\text{expit}(\alpha_2^*)+[1-\text{expit}(\alpha_2^*)]\exp[-\exp(\alpha_1^*+o_n)]\}\phi_2(\valpha^*;\vmu_{\backslash n}^*,\mSigma_{\backslash n}^*) & \text{if }y_n=0, \\
        \{[1-\text{expit}(\alpha_2^*)]\exp[y_n(\alpha_1^*+o_n)-\exp(\alpha_1^*+o_n)]/y_n!\}\phi_2(\valpha^*;\vmu_{\backslash n}^*,\mSigma_{\backslash n}^*) & \text{if }y_n>0.
    \end{cases}
\end{align*}
There is no way to further reduce the integrals $I_{h_{f,n}^*}^0$, $\vI_{h_{f,n}^*}^1$, and $\mI_{h_{f,n}^*}^2$; instead, they may be computed using bivariate numerical quadrature.

\subsection{Binomial regression}

For a refinement of a likelihood site approximation $\widetilde{f}_n(\vtheta)$, the kernel of the low-dimensional tilted distribution is given by
\begin{align*}
    \widetilde{h}_{f,n}^*(\alpha^*)&=f^*(\alpha^*,y_n,\aleph_n)\phi(\alpha^*;\mu_{\backslash n}^*,\Sigma_{\backslash n}^*) \\
    &=\binom{\aleph_n}{y_n}\Phi(\alpha^*)^{y_n}[1-\Phi(\alpha^*)]^{\aleph_n-y_n}\phi(\alpha^*;\mu_{\backslash n}^*,\Sigma_{\backslash n}^*).
\end{align*}
The integrals $I_{h_{f,n}^*}^0$, $I_{h_{f,n}^*}^1$, and $I_{h_{f,n}^*}^2$ may then be computed straightforwardly using univariate numerical quadrature.

\section{Dataset descriptions}\label{sec:data}

Here, we describe and reference the real-world datasets used in Section \ref{sec:experiments}.

\subsection{Zero-inflated Poisson regression}

\texttt{Epilepsy} \citep{thall1990some} contains data from a clinical trial of epileptic participants.
The response is the number of seizures, which is predicted by covariates that include treatment and week of visit.
Each group in \texttt{Epilepsy} corresponds to a participant.
\texttt{Fly} was a dataset obtained from a collaborator, and stores the number of eggs laid (the response) corresponding to predictors such as cohort and diet in a nutrition experiment on Canton flies.
Each group in \texttt{Fly} corresponds to a fly.
Both \texttt{Owl\_1} and \texttt{Owl\_3} are from the same dataset \citep{roulin2007nestling} which corresponds to a study of begging behaviour in infant barn owls; the response is nestling vocalisation in the absence of parents, which are predicted by variables such as the sex of the provisioning parent and the arrival time of the parent.
Each group corresponds to a nest.
For \texttt{Owl\_1}, we only assumed a random intercept across groups, while for \texttt{Owl\_3}, we also assumed random slopes for parent arrival time and its square---$P$ decreases by two for \texttt{Owl\_3} when compared to \texttt{Owl\_1} because we also removed, in addition to parent arrival time, an interaction term involving parent arrival time from the predictors with fixed effects.
\texttt{Salamanders} \citep{price2016effects,price2016data} contains repeated site data from an ecological survey, where the response is the number of salamanders observed and the predictors include variables such as mining impact and days since precipitation.
Each group in \texttt{Salamanders} corresponds to a site, and we assumed random slopes for water temperature and its square, and days since precipitation.
Finally, \texttt{Webworms} \citep{beall1940fit} contains counts of webworms in a beetroot field, with these counts being predicted by factors such as presence of spray and/or lead treatment.
Each group in \texttt{Webworms} corresponds to a block in the field, and we assumed random slopes for the presence of lead treatment.

\subsection{Binomial regression}

\texttt{CTSIB} \citep{steele1998effect} contains balance measurements (the response) which can be converted into a binary variable, along with predictors such as age and vision, for study measuring the effect of surface and vision on balance on a large number of participants.
Each group in \texttt{CTSIB} corresponds to a participant.
Both \texttt{HDP\_1} and \texttt{HDP\_3} are from the same dataset \citep{hdp}, which stores data on remission status (the response), in addition to predictors such as hospital and doctor, for a large number of cancer patients.
The dataset may be aggregated to a doctor level, where the response is now binomial (number of patients with remission), and the predictors are doctor-specific.
We used this dataset even though it technically simulated, as it is a very popular benchmark.
Each group corresponds to a hospital.
For \texttt{HDP\_1}, we only assumed a random intercept across groups, while for \texttt{HDP\_3}, we also assumed random slopes for the number of lawsuits and level of education.
\texttt{Salamanders} is the same dataset used for the experiments for the ZIP model, except the original count response is now converted into a binary variable indicating presence/absence.
\texttt{Tapped} \citep{kilbourn2017speech} contains observations on whether or not a letter was pronounced as a `tap' (the response) in a speech production experiment, with associated predictors including speech rate and whether or not the letter was located before a syntactic juncture.
Each group in \texttt{Tapped} corresponds to a participant, and we assumed random slopes for vowel duration.
Finally, \texttt{Toenail} \citep{de1998twelve} records toenail infection status (the response) for the participants of a toenail infection treatment study, and also includes predictors such as treatment and month of visit.
Each group in \texttt{Toenail} corresponds to a participant.

For the NLSY dataset, we considered the following years as they had the most data: 2002, 2007, 2009, 2010, 2011, 2015, 2017.
Among these years, we used a combination of year-varying year-constant covariates/survey questions to predict the response to the health status survey question (reference codes S1225000, T1049500, T4562200, T6206400, T7703800, U1096500, and U2962600).
These were: birth year (reference code R0536402), sex (reference code R0536300), ethnicity (reference code R1482600), age of mother when born (reference code R1200200), father's highest grade (reference code R1302400), mother's highest grade (reference code R1302500), health of one parent (reference code R0607900), whether one parent's health limits their employment (reference code R0608000), urban/rural residence (reference code R1217500), percentage of peers who get drunk (reference code R0070500), percentage of peers who smoke (reference code R0070400), percentage of peers who use illegal drugs (reference code R0071000), anxious self-rating (reference code T3162503), critical self-rating (reference code T3162501), dependable self-rating (reference code T3162502), extroverted self-rating (reference code T3162500), relationship with God (reference code S0919700), prays more than once a day (reference code S0919800), rating of current life (reference code T3162400), year, hours of sleep per night (reference codes S1225600, T1050100, T4565600, T6209800, T7707200, U1099900, and U2963000), and income (reference codes S1055800, T0889800, T4406000, T6055500, T7545600, U0956900, and U2857200).
We also considered the following interactions between covariates/survey questions in the prediction of health status: interaction of ethnicity and peer covariates, interaction of ethnicity and personality covariates, and two-way interactions of personality covariates.
The effect of hours of sleep per night and income on health status was assumed to vary between participants (i.e., random effects were assumed for these covariates).

\section{Additional results}\label{sec:more_results}

We provide additional results for the experiments conducted in Section \ref{sec:experiments}; these aim to complement the results shown in the main text.
The additional results can be split up into marginal deviations, marginal distribution plots, and maximum mean discrepancies.

\subsection{Marginal deviations}

Table \ref{table:zip_marginal} and Table \ref{table:binom_marginal} show the marginal deviations (non-absolute) corresponding to the second set of experiments for the ZIP and binomial models respectively.
A positive/negative value of $\overline{\text{dev}(\widehat{\mu}_i)}$ indicates over/underestimation of the marginal posterior means, while a value of $\overline{\text{dev}(\widehat{\sigma}_i)}$ above/below one indicates over/underestimation of the marginal posterior SD.
Perhaps the main trend for both model types is that all ABI methods tended to underestimate the marginal posterior SD across all components, with this being most noticeable in the components of $\text{vech}(\mSigma)$.

\begin{table}
\centering
\fontsize{9pt}{9pt}\selectfont
\begin{tabular}{@{}cccccccc@{}}
\toprule
\multirow{2}{*}{Dataset}     & \multirow{2}{*}{Method} & \multicolumn{5}{c}{$\overline{\text{dev}(\widehat{\mu}_i)}$/$\overline{\text{dev}(\widehat{\sigma}_i)}$} \\
                             &                         & $(\vtheta,\text{vech}(\mSigma))$       & $\vu$       & $\lambda$      & $\vbeta$      & $\text{vech}(\mSigma)$                             \\ \midrule
\multirow{3}{*}{\texttt{Epilepsy}}         & EP-S                      & 0.04/1.00 & 0.04/1.01 & 0.44/0.75 & -0.02/1.00 &  0.00/0.93              \\
                             & INLA                    & 0.05/0.98 & 0.06/0.98 & 0.18/0.93 &  0.00/0.99 & -0.07/0.94              \\
                             & GVB                     & 0.04/0.99 & 0.04/0.99 & 0.10/0.98 &  0.03/0.99 & -0.14/0.91               \\ \midrule
\multirow{3}{*}{\texttt{Fly}} & EP-S                      & -0.02/0.99 & -0.02/1.00 &  0.02/0.96 &  0.01/0.84 &  0.09/0.84               \\
                             & INLA                    & 0.05/0.99 &  0.05/0.99 &  0.04/0.92 & -0.01/0.89 & -0.05/0.96               \\
                             & GVB                     & 0.05/0.92 &  0.05/0.93 & -0.35/1.02 &  0.02/0.55 &  0.18/0.96               \\ \midrule
\multirow{3}{*}{\texttt{Owl\_1}} & EP-S                      & 0.00/1.01 & 0.00/1.02 & -0.02/0.99 & 0.00/0.98 &  0.18/0.85              \\
                             & INLA                    & 0.02/0.97 & 0.03/0.97 &  0.07/0.91 & 0.00/1.00 & -0.15/0.91               \\
                             & GVB                     & 0.01/0.95 & 0.01/0.94 & -0.04/1.01 & 0.09/0.98 & -0.29/0.78               \\ \midrule
\multirow{3}{*}{\texttt{Owl\_3}}         & EP-S                      & -0.01/0.99 & -0.02/1.01 & -0.15/1.02 & -0.02/1.00 &  0.10/0.82             \\
                             & INLA                    &  0.02/0.97 &  0.03/0.97 &  0.07/0.96 & -0.03/0.99 & -0.03/0.93              \\
                             & GVB                     & -0.02/0.91 &  0.00/0.94 &  0.18/0.96 &  0.00/0.95 & -0.39/0.62              \\ \midrule
\multirow{3}{*}{\texttt{Salamanders}} & EP-S                      & 0.02/0.99 & 0.02/1.03 & -0.06/1.00 & -0.16/1.02 &  0.05/0.69              \\
                             & INLA                    & 0.03/0.95 & 0.05/0.97 & -0.01/0.99 & -0.10/1.06 & -0.06/0.77              \\
                             & GVB                     & 0.00/0.73 & 0.01/0.79 & -0.01/1.01 &  0.11/0.74 & -0.15/0.33              \\ \midrule
\multirow{3}{*}{\texttt{Webworms}}    & EP-S                      & -0.02/0.95 & -0.03/0.96 & 0.06/0.92 &  0.03/0.93 &  0.02/0.90              \\
                             & INLA                    &  0.01/0.96 &  0.03/0.98 & 0.05/0.98 & -0.05/0.96 & -0.11/0.77              \\
                             & GVB                     & -0.01/0.89 &  0.03/0.93 & 0.08/0.92 & -0.07/0.95 & -0.32/0.62              \\ \bottomrule
\end{tabular}
\caption{Marginal deviations from MCMC of ABI methods when performing inference on a Bayesian mixed-effects ZIP regression model, for smaller to medium-sized datasets.}
\label{table:zip_marginal}
\end{table}

\begin{table}
\centering
\fontsize{9pt}{9pt}\selectfont
\begin{tabular}{@{}cccccccc@{}}
\toprule
\multirow{2}{*}{Dataset}     & \multirow{2}{*}{Method} & \multicolumn{4}{c}{$\overline{\text{dev}(\widehat{\mu}_i)}$/$\overline{\text{dev}(\widehat{\sigma}_i)}$}  \\
                             &                         & $(\vtheta,\text{vech}(\mSigma))$       & $\vu$ & $\vbeta$      & $\text{vech}(\mSigma)$                                 \\ \midrule
\multirow{3}{*}{\texttt{CTSIB}}         & EP-S                      &  0.02/0.96 &  0.02/0.99 &  0.02/0.90 &  0.12/0.56              \\
                             & INLA                    & -0.05/0.90 & -0.02/0.92 & -0.18/0.87 & -0.23/0.48              \\
                             & GVB                     & -0.01/0.87 &  0.01/0.89 & -0.05/0.84 & -0.44/0.56              \\ \midrule
\multirow{3}{*}{\texttt{HDP\_1}} & EP-S                      &  0.03/1.00 &  0.04/1.00 & -0.02/1.01 &  0.04/0.95              \\
                             & INLA                    &  0.02/1.00 &  0.02/1.00 & -0.02/0.98 & -0.06/0.97              \\
                             & GVB                     & -0.02/0.98 & -0.02/0.99 &  0.03/0.97 & -0.25/0.85              \\ \midrule
\multirow{3}{*}{\texttt{HDP\_3}} & EP-S                      & 0.01/1.00 & 0.00/1.00 & -0.02/1.00 &  0.03/0.88              \\
                             & INLA                    & 0.01/0.98 & 0.01/0.99 & -0.01/1.00 & -0.03/0.94              \\
                             & GVB                     & 0.06/0.95 & 0.00/0.96 & -0.16/0.93 &  1.30/0.74              \\ \midrule
\multirow{3}{*}{\texttt{Salamanders}}         & EP-S                      &  0.00/0.98 &  0.00/1.02 & -0.10/1.01 &  0.02/0.67              \\
                             & INLA                    &  0.00/0.94 &  0.00/0.96 & -0.01/0.97 & -0.08/0.78              \\
                             & GVB                     & -0.02/0.76 & -0.01/0.82 &  0.08/0.79 & -0.14/0.38              \\ \midrule
\multirow{3}{*}{\texttt{Tapped}} & EP-S                      &  0.00/0.94 &  0.01/0.97 &  0.02/0.98 & -0.03/0.63              \\
                             & INLA                    & -0.02/0.95 & -0.01/0.95 & -0.02/0.97 & -0.07/0.80              \\
                             & GVB                     & -0.02/0.85 &  0.00/0.88 & -0.03/0.93 & -0.26/0.51              \\ \midrule
\multirow{3}{*}{\texttt{Toenail}}    & EP-S                      & -0.04/0.88 & -0.04/0.88 & 0.18/0.88 & -0.89/0.36              \\
                             & INLA                    & -0.17/0.82 & -0.19/0.82 & 1.12/0.86 & -1.46/0.61              \\
                             & GVB                     &  0.04/0.90 &  0.04/0.90 & 0.04/0.85 & -0.75/0.46              \\ \bottomrule
\end{tabular}
\caption{Marginal deviations from MCMC of ABI methods when performing inference on a Bayesian mixed-effects binomial regression model, for smaller to medium-sized datasets.}
\label{table:binom_marginal}
\end{table}

Table \ref{table:binom_marginal_big} shows the marginal deviations (non-absolute) corresponding to the binomial experiments on the NLSY dataset.
When comparing EP-S to INLA, there seems to be a slight overestimation in the marginal posterior SD of the $\vu$ and $\vbeta$ components, and an overestimation and underestimation in the marginal posterior mean and SD respectively of the components of $\text{vech}(\mSigma)$.

\begin{table}
\centering
\fontsize{9pt}{9pt}\selectfont
\begin{tabular}{@{}cccccccc@{}}
\toprule
\multicolumn{4}{c}{$\overline{\text{dev}(\widehat{\mu}_i)}$/$\overline{\text{dev}(\widehat{\sigma}_i)}$} \\
                            $(\vtheta,\text{vech}(\mSigma))$       & $\vu$ & $\vbeta$      & $\text{vech}(\mSigma)$      &                           \\ \midrule
                             0.01/1.23 & 0.01/1.23 & 0.01/1.06 & 1.98/0.28             \\ \bottomrule
\end{tabular}
\caption{Marginal deviations from INLA of distributed EP-S when performing inference on a Bayesian mixed-effects binomial regression model for the NLSY dataset.}
\label{table:binom_marginal_big}
\end{table}

\subsection{Marginal distribution plots}

\begin{figure}
    \begin{center}
        \includegraphics[scale=0.9]{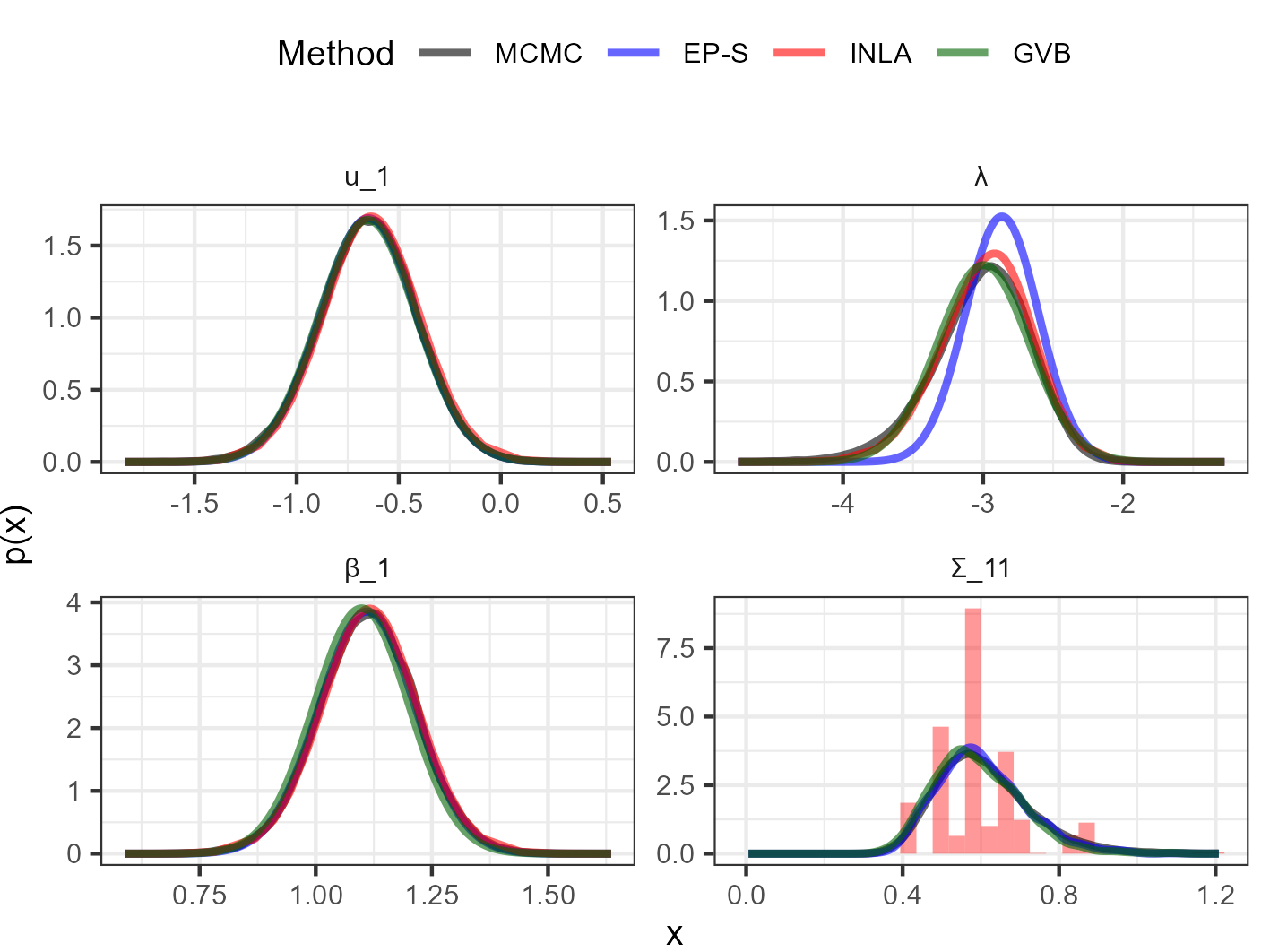}
    \end{center}
    \caption{Bayesian mixed-effects ZIP regression model, fit to \texttt{Epilepsy}---marginal posterior distributions for $u_1$, $\kappa$, $\lambda$, and $\beta_1$ (estimated using MCMC) and their approximations using ABI methods.}
    \label{fig:zip_epilepsy}
\end{figure}

\begin{figure}
    \begin{center}
        \includegraphics[scale=0.9]{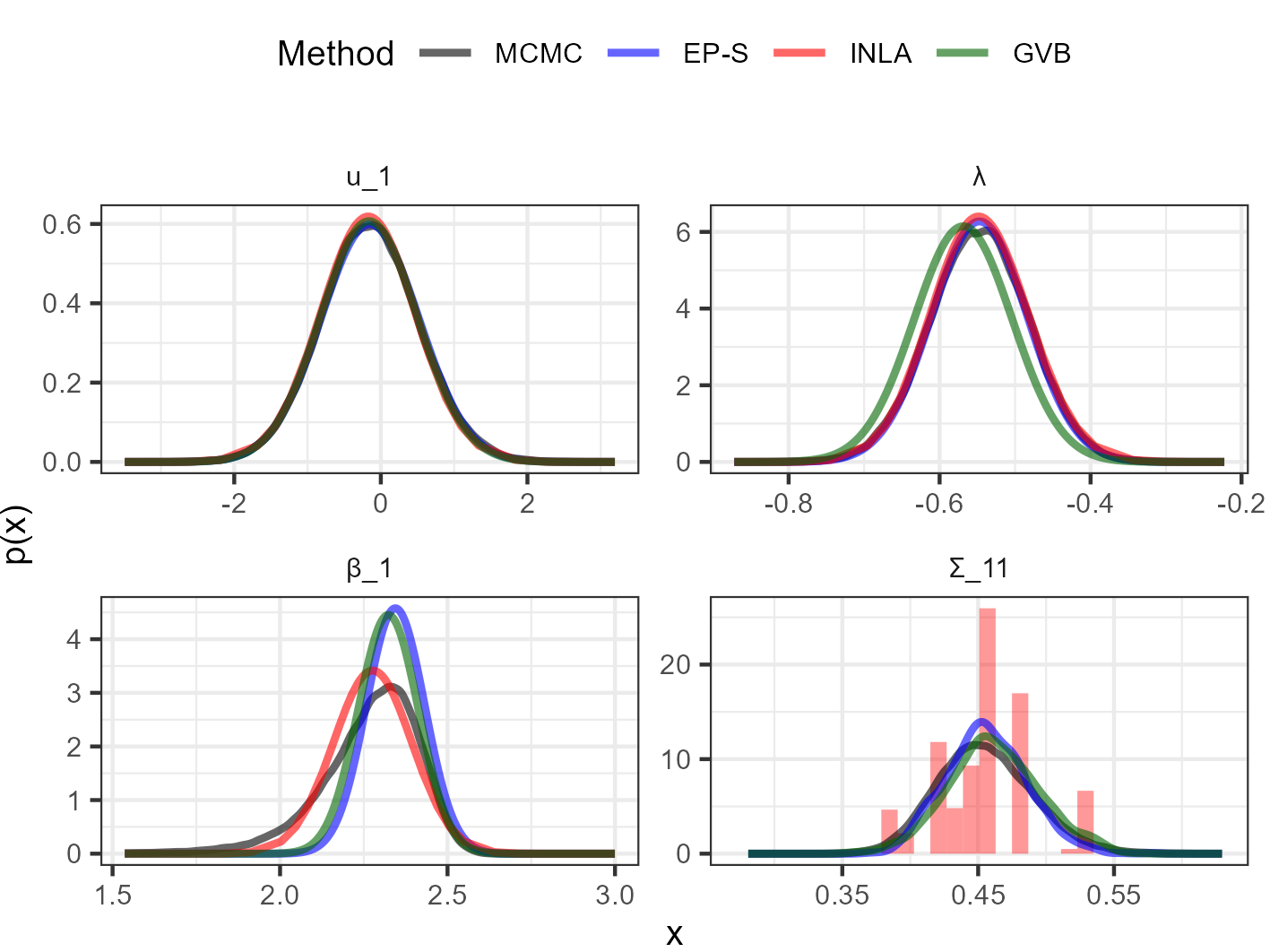}
    \end{center}
    \caption{Bayesian mixed-effects ZIP regression model, fit to \texttt{Fly}---marginal posterior distributions for $u_1$, $\kappa$, $\lambda$, and $\beta_1$ (estimated using MCMC) and their approximations using ABI methods.}
    \label{fig:zip_fly}
\end{figure}

\begin{figure}
    \begin{center}
        \includegraphics[scale=0.9]{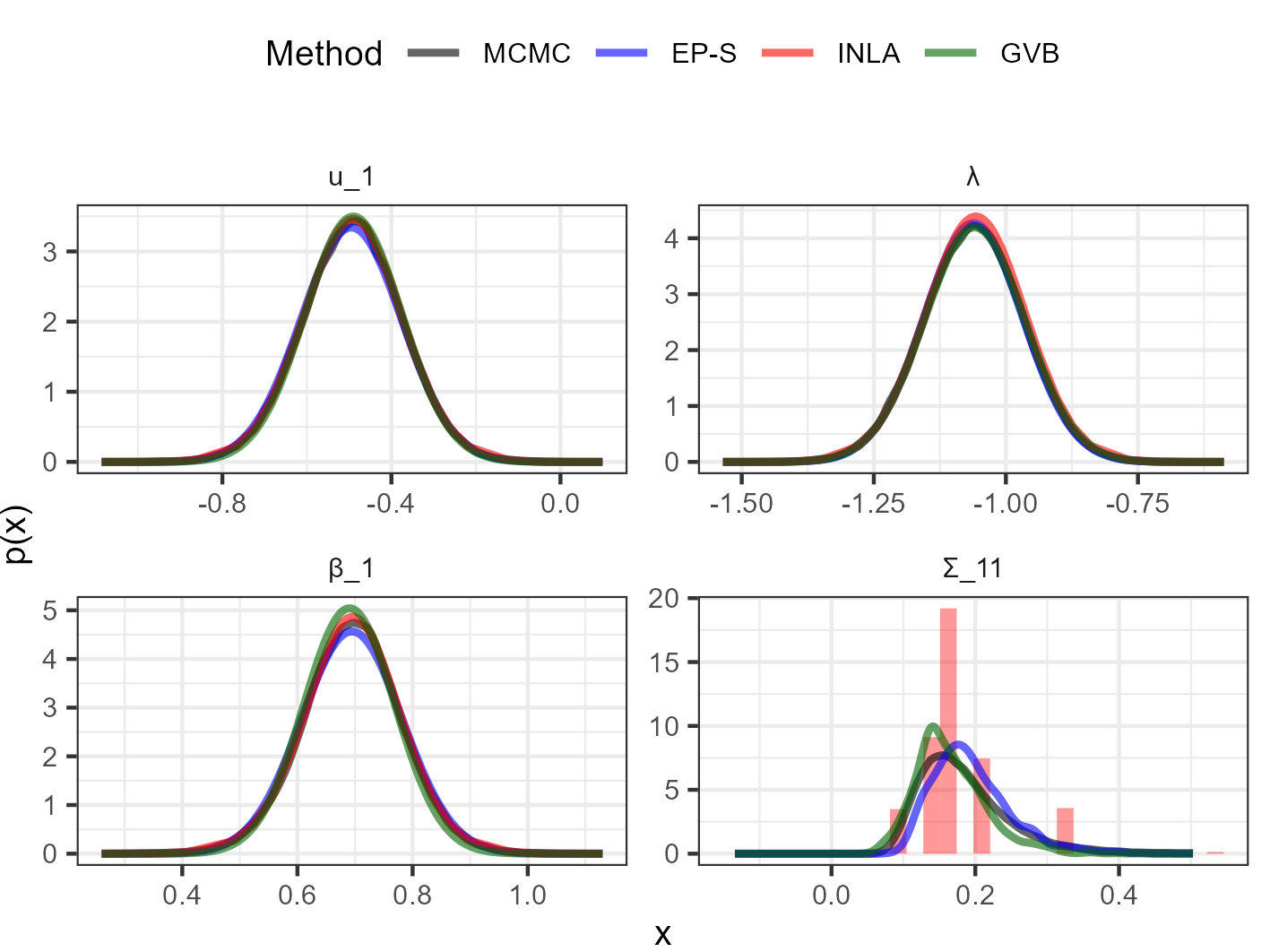}
    \end{center}
    \caption{Bayesian mixed-effects ZIP regression model, fit to \texttt{Owl\_1}---marginal posterior distributions for $u_1$, $\kappa$, $\lambda$, and $\beta_1$ (estimated using MCMC) and their approximations using ABI methods.}
    \label{fig:zip_owl_1}
\end{figure}

\begin{figure}
    \begin{center}
        \includegraphics[scale=0.9]{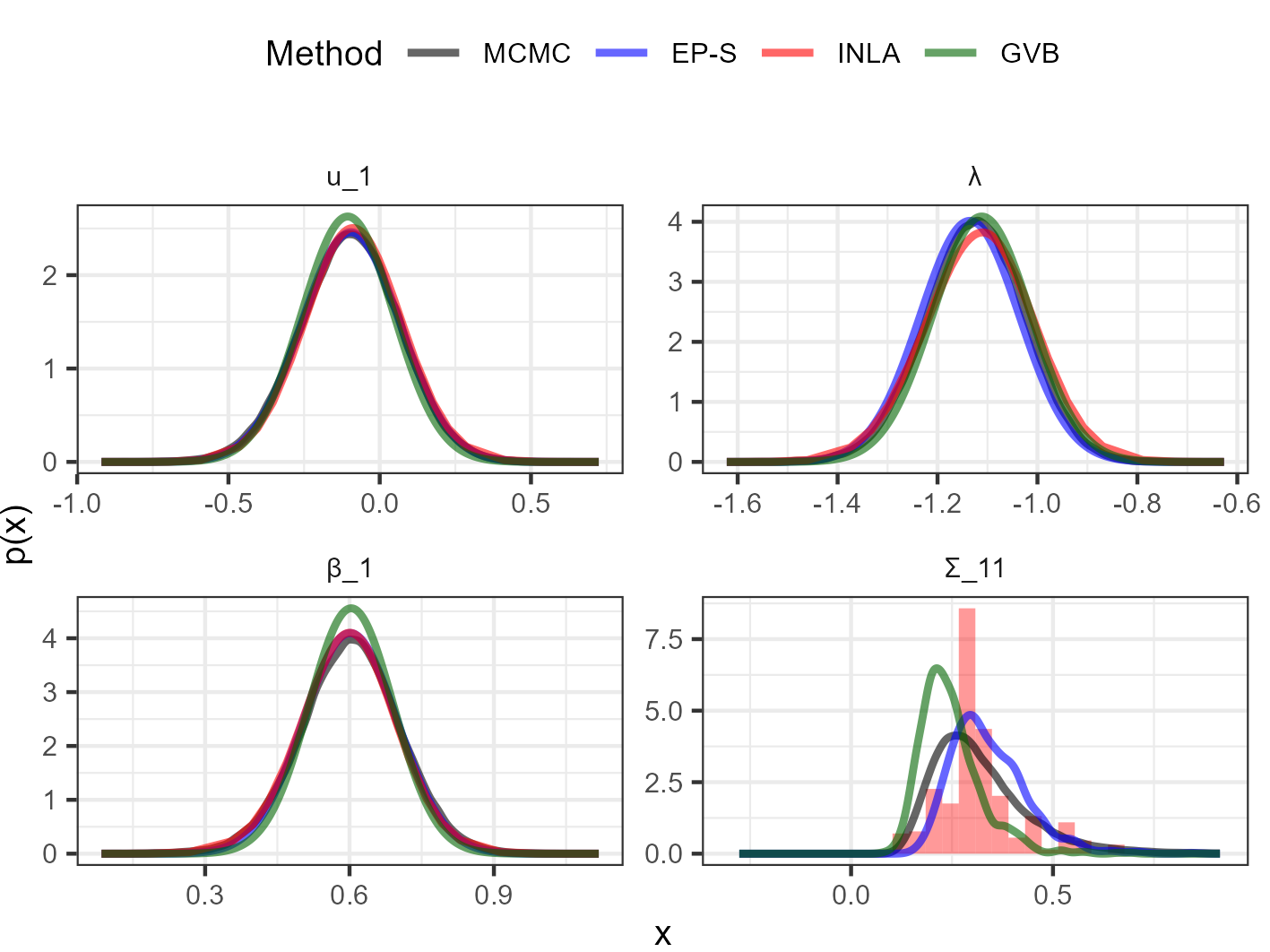}
    \end{center}
    \caption{Bayesian mixed-effects ZIP regression model, fit to \texttt{Owl\_3}---marginal posterior distributions for $u_1$, $\lambda$, $\beta_1$, and $\Sigma_{1,1}$ (estimated using MCMC) and their approximations using ABI methods.}
    \label{fig:zip_owl_3}
\end{figure}

\begin{figure}
    \begin{center}
        \includegraphics[scale=0.9]{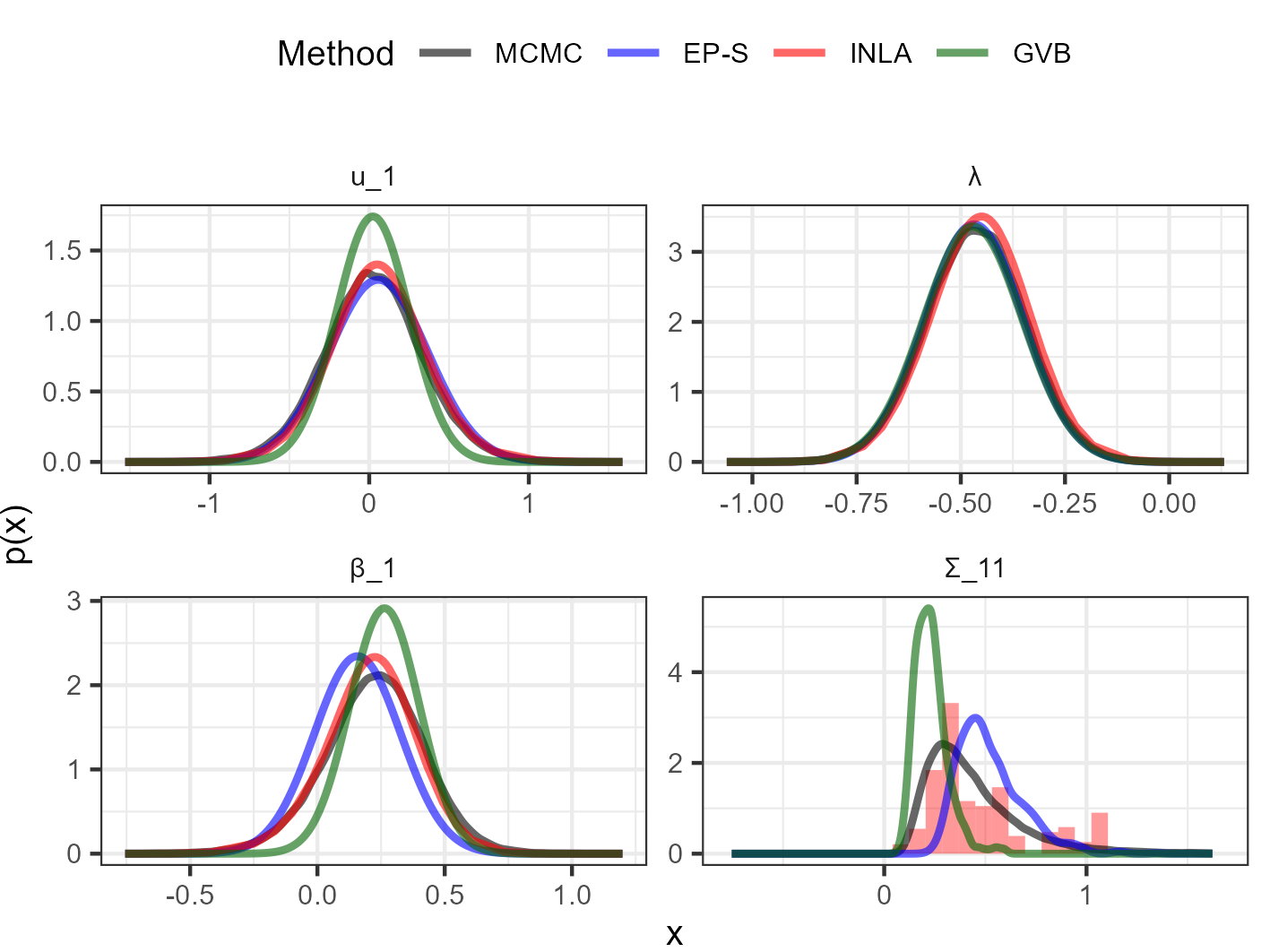}
    \end{center}
    \caption{Bayesian mixed-effects ZIP regression model, fit to \texttt{Salamanders}---marginal posterior distributions for $u_1$, $\lambda$, $\beta_1$, and $\Sigma_{1,1}$ (estimated using MCMC) and their approximations using ABI methods.}
    \label{fig:zip_salamanders}
\end{figure}

\begin{figure}
    \begin{center}
        \includegraphics[scale=0.9]{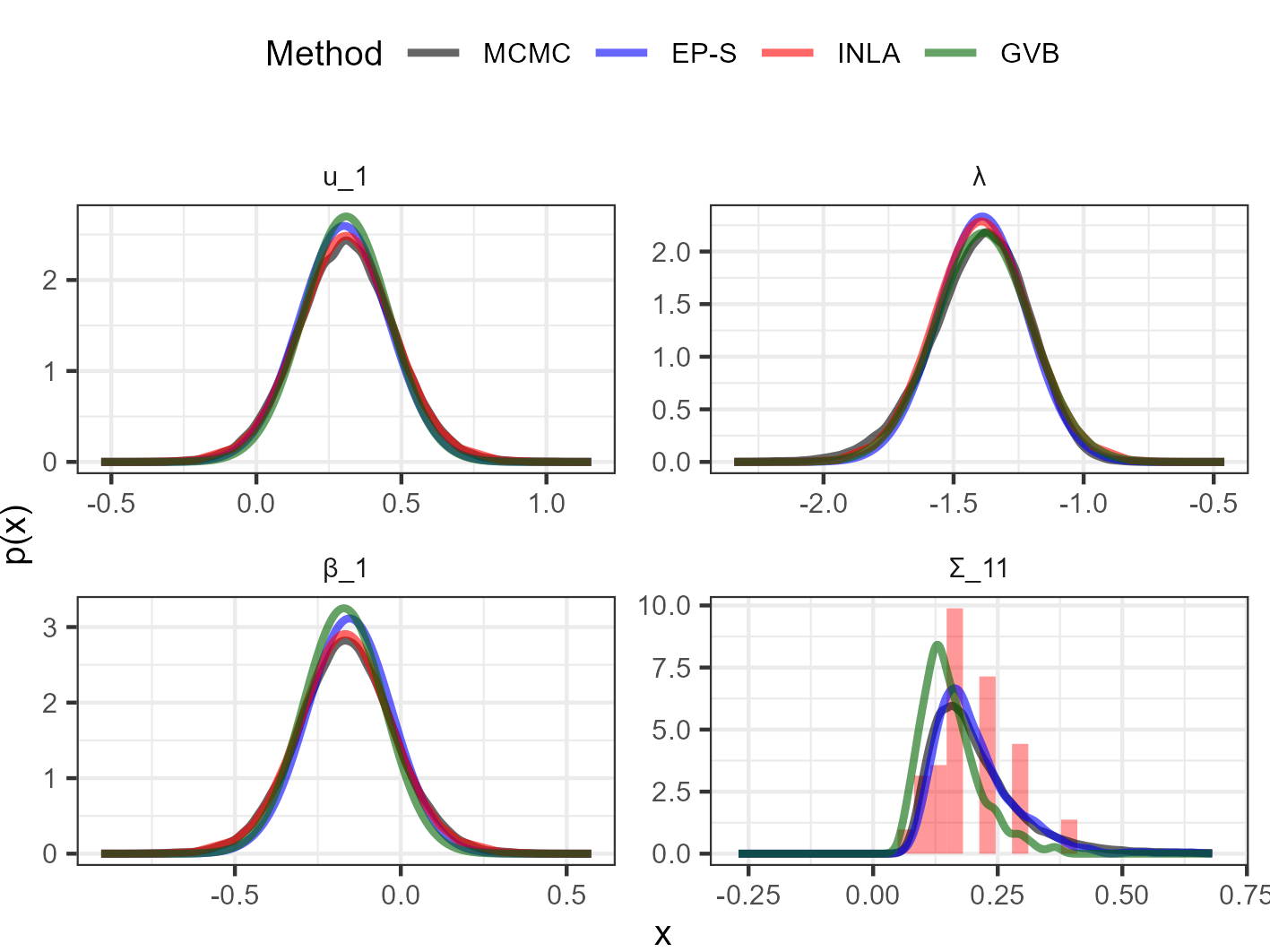}
    \end{center}
    \caption{Bayesian mixed-effects ZIP regression model, fit to \texttt{Webworms}---marginal posterior distributions for $u_1$, $\lambda$, $\beta_1$, and $\Sigma_{1,1}$ (estimated using MCMC) and their approximations using ABI methods.}
    \label{fig:zip_webworms}
\end{figure}

\begin{figure}
    \begin{center}
        \includegraphics[scale=0.9]{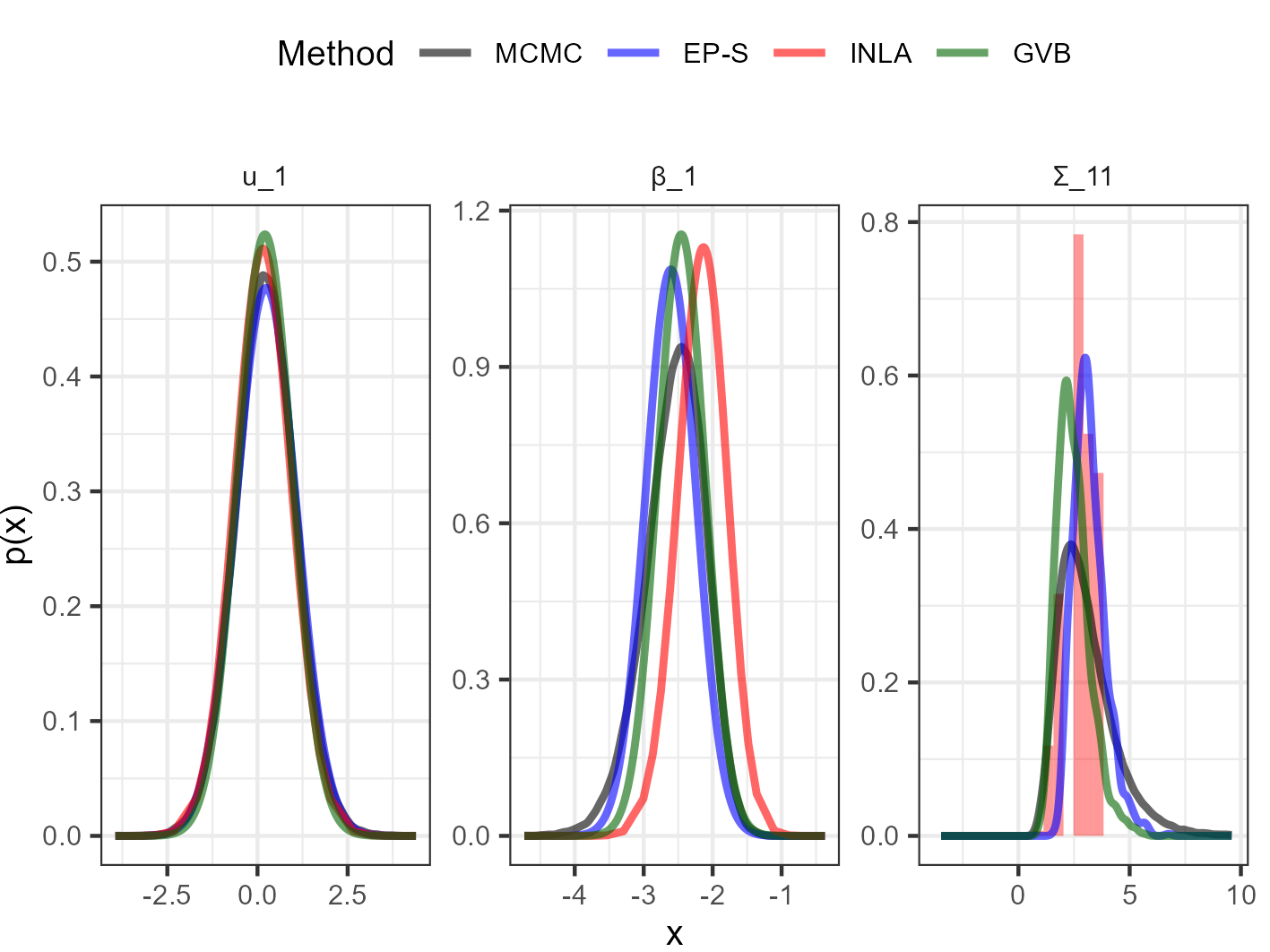}
    \end{center}
    \caption{Bayesian mixed-effects binomial regression model, fit to \texttt{CTSIB}---marginal posterior distributions for $u_1$, $\kappa$, and $\beta_1$ (estimated using MCMC) and their approximations using ABI methods.}
    \label{fig:binom_ctsib}
\end{figure}

\begin{figure}
    \begin{center}
        \includegraphics[scale=0.9]{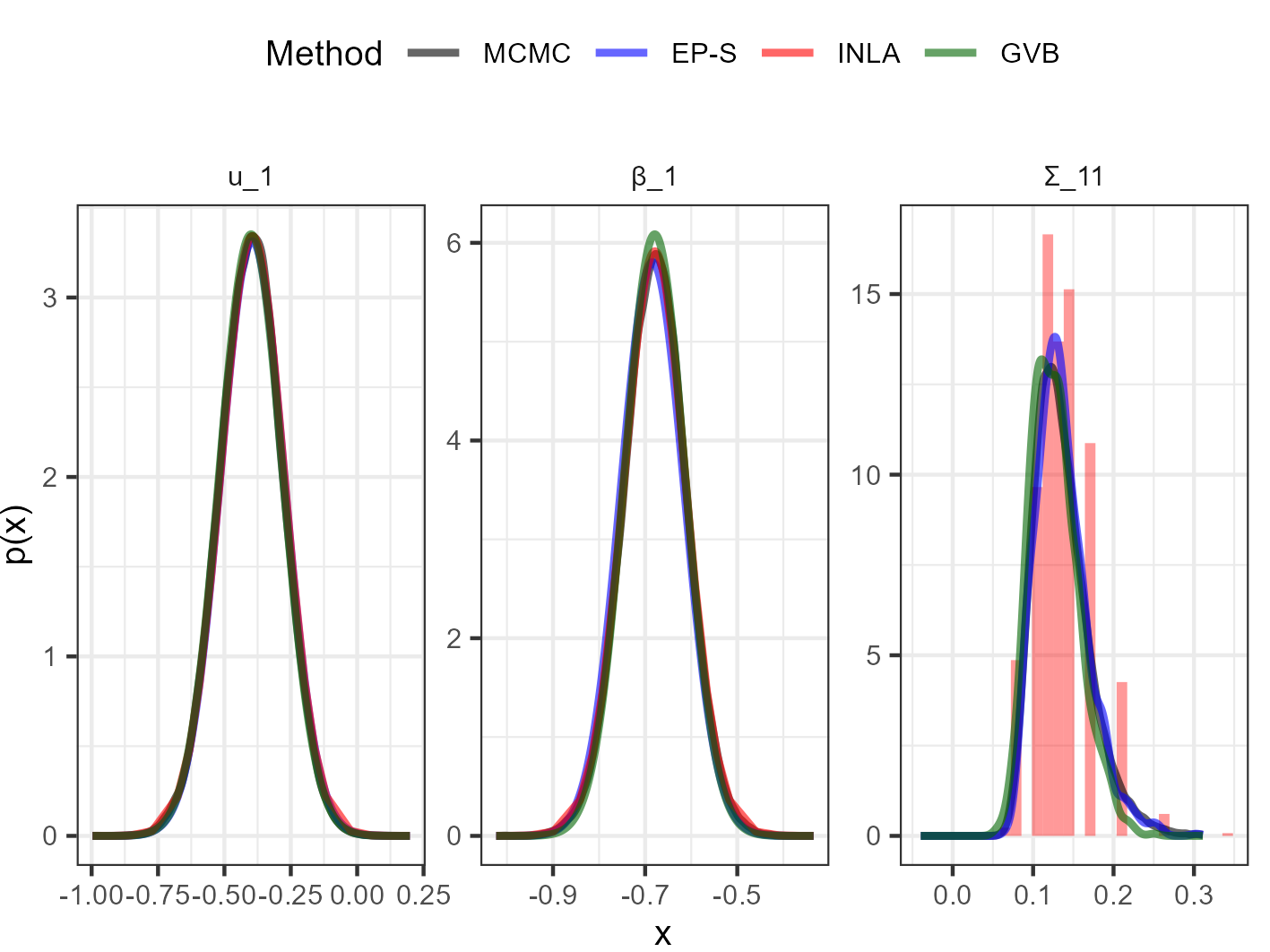}
    \end{center}
    \caption{Bayesian mixed-effects binomial regression model, fit to \texttt{HDP\_1}---marginal posterior distributions for $u_1$, $\kappa$, and $\beta_1$ (estimated using MCMC) and their approximations using ABI methods.}
    \label{fig:binom_hdp_1}
\end{figure}

\begin{figure}
    \begin{center}
        \includegraphics[scale=0.9]{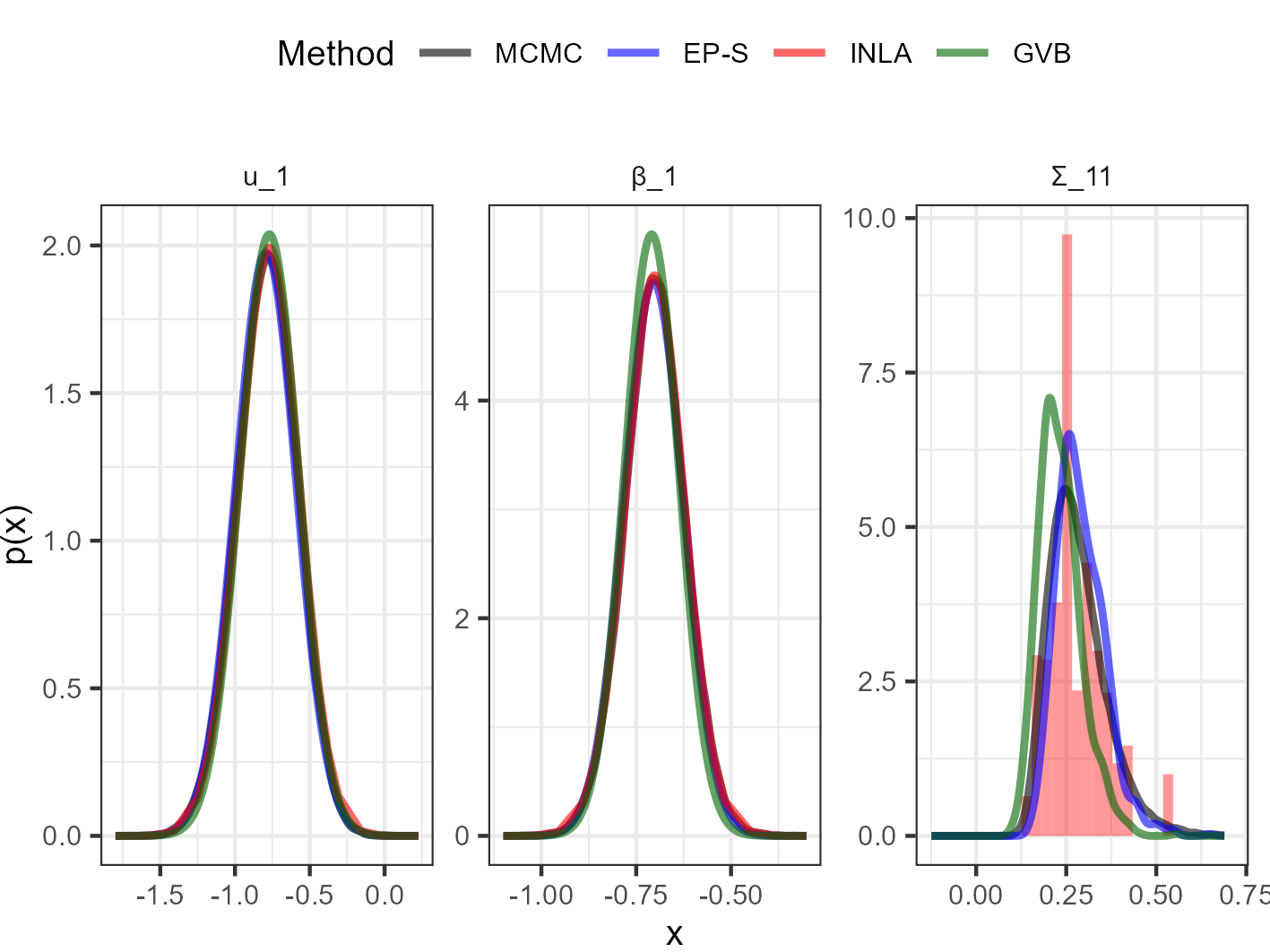}
    \end{center}
    \caption{Bayesian mixed-effects binomial regression model, fit to \texttt{HDP\_3}---marginal posterior distributions for $u_1$, $\kappa$, and $\beta_1$ (estimated using MCMC) and their approximations using ABI methods.}
    \label{fig:binom_hdp_3}
\end{figure}

\begin{figure}
    \begin{center}
        \includegraphics[scale=0.9]{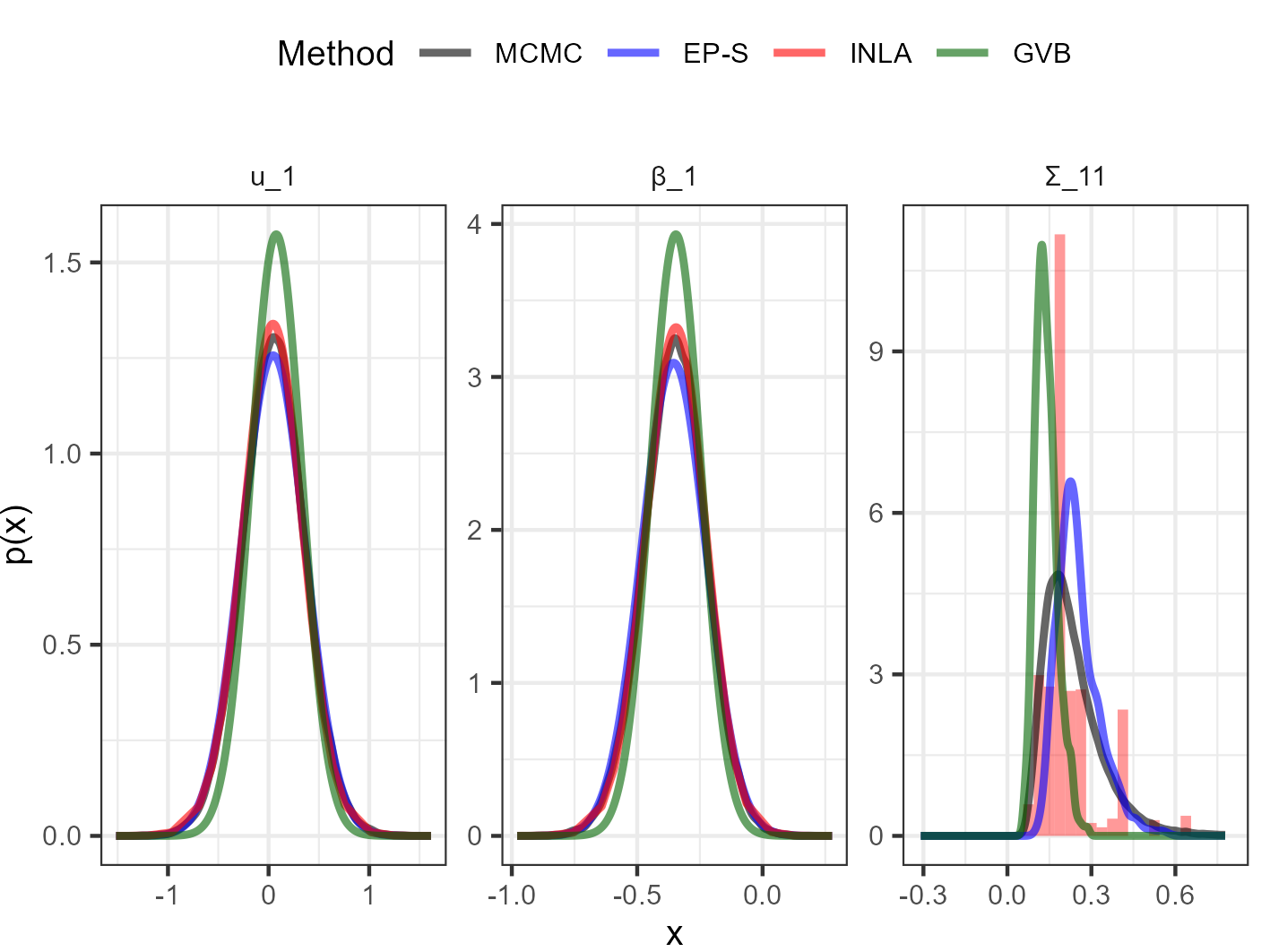}
    \end{center}
    \caption{Bayesian mixed-effects binomial regression model, fit to \texttt{Salamanders}---marginal posterior distributions for $u_1$, $\beta_1$, and $\Sigma_{1,1}$ (estimated using MCMC) and their approximations using ABI methods.}
    \label{fig:binom_salamanders}
\end{figure}

\begin{figure}
    \begin{center}
        \includegraphics[scale=0.9]{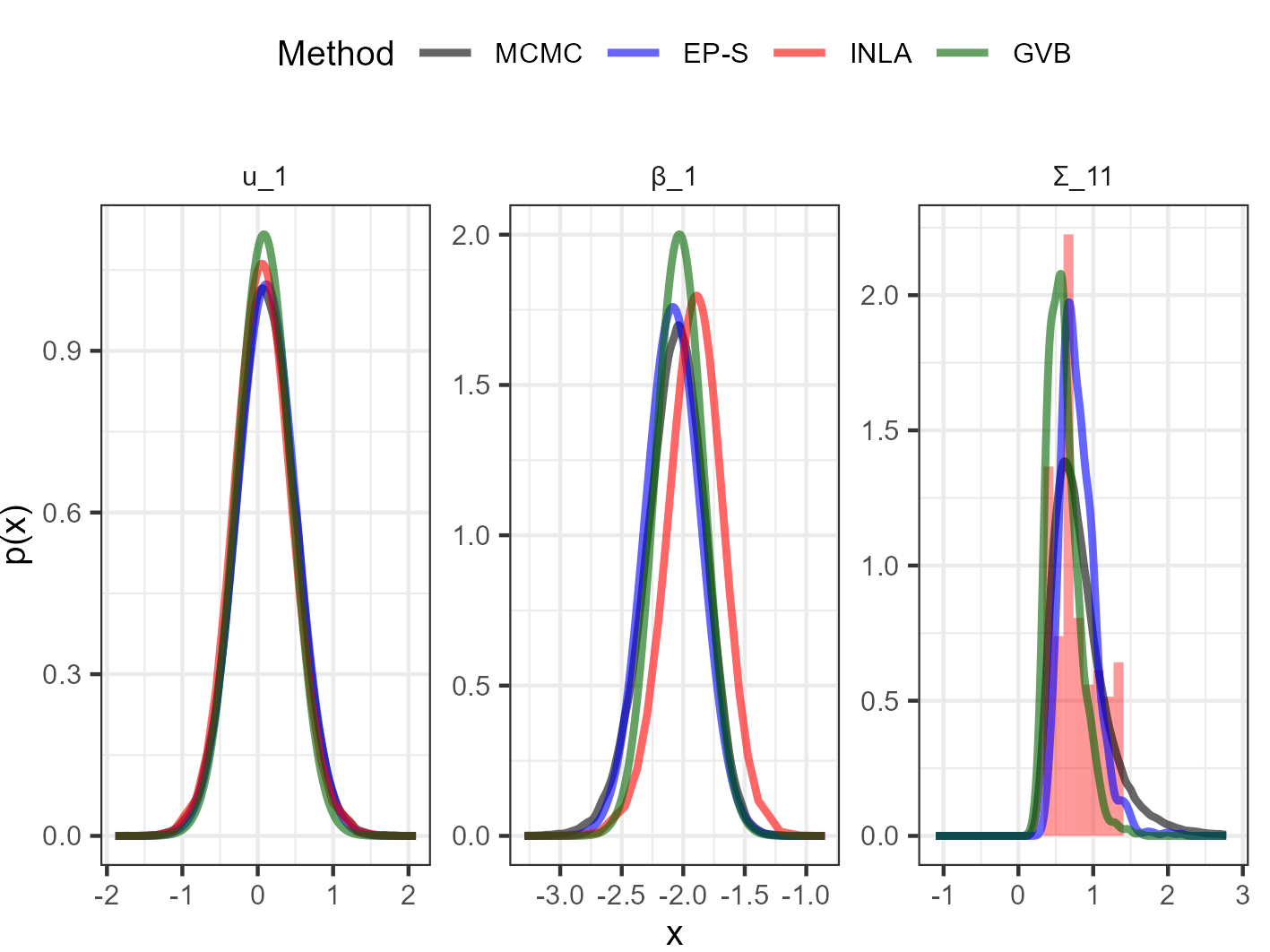}
    \end{center}
    \caption{Bayesian mixed-effects binomial regression model, fit to \texttt{Tapped}---marginal posterior distributions for $u_1$, $\beta_1$, and $\Sigma_{1,1}$ (estimated using MCMC) and their approximations using ABI methods.}
    \label{fig:binom_tapped}
\end{figure}

\begin{figure}
    \begin{center}
        \includegraphics[scale=0.9]{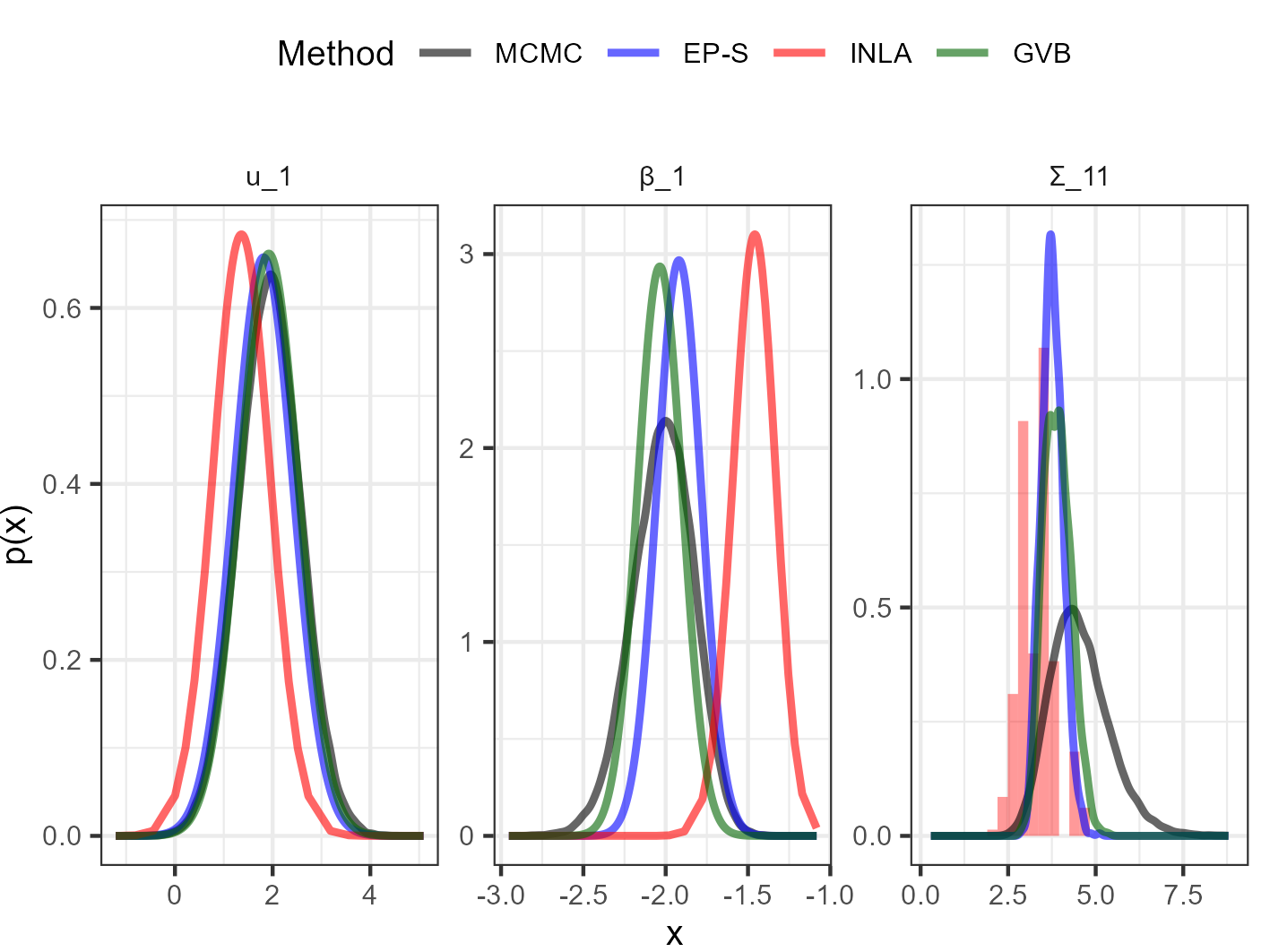}
    \end{center}
    \caption{Bayesian mixed-effects binomial regression model, fit to \texttt{Toenail}---marginal posterior distributions for $u_1$, $\beta_1$, and $\Sigma_{1,1}$ (estimated using MCMC) and their approximations using ABI methods.}
    \label{fig:binom_toenail}
\end{figure}

Marginal distribution plots for the second set of ZIP model experiments are given from Figure \ref{fig:zip_epilepsy} to Figure \ref{fig:zip_webworms}, while the same plots for the second set of binomial model experiments are given from Figure \ref{fig:binom_ctsib} to Figure \ref{fig:binom_toenail}.
For all methods, the marginal distributions associated with the components of $\mSigma$ were estimated from the samples of the posterior approximation, using Silverman's rule of thumb for bandwidth selection \citep{silverman2018density}; this was because the parameterisation of the random-effects covariance matrix in INLA did not match ours.
As the INLA samples of the components of $\mSigma$ were very discrete, we made the decision to represent the marginal distribution in this case using a density histogram, rather than a density curve.
Finally, for the sake of conciseness, we do not plot all marginal distributions, but rather only the first of each type of marginal distribution; for example, we only plot the marginal distribution of $u_1$ instead of all components of $\vu$.
While there was variation in the accuracy of EP-S, INLA, and GVB within the different types of marginal distributions, the first marginal distribution was generally representative of the rest.
These marginal distribution plots generally agree with the results from the main text.

\begin{figure}
    \begin{center}
        \includegraphics{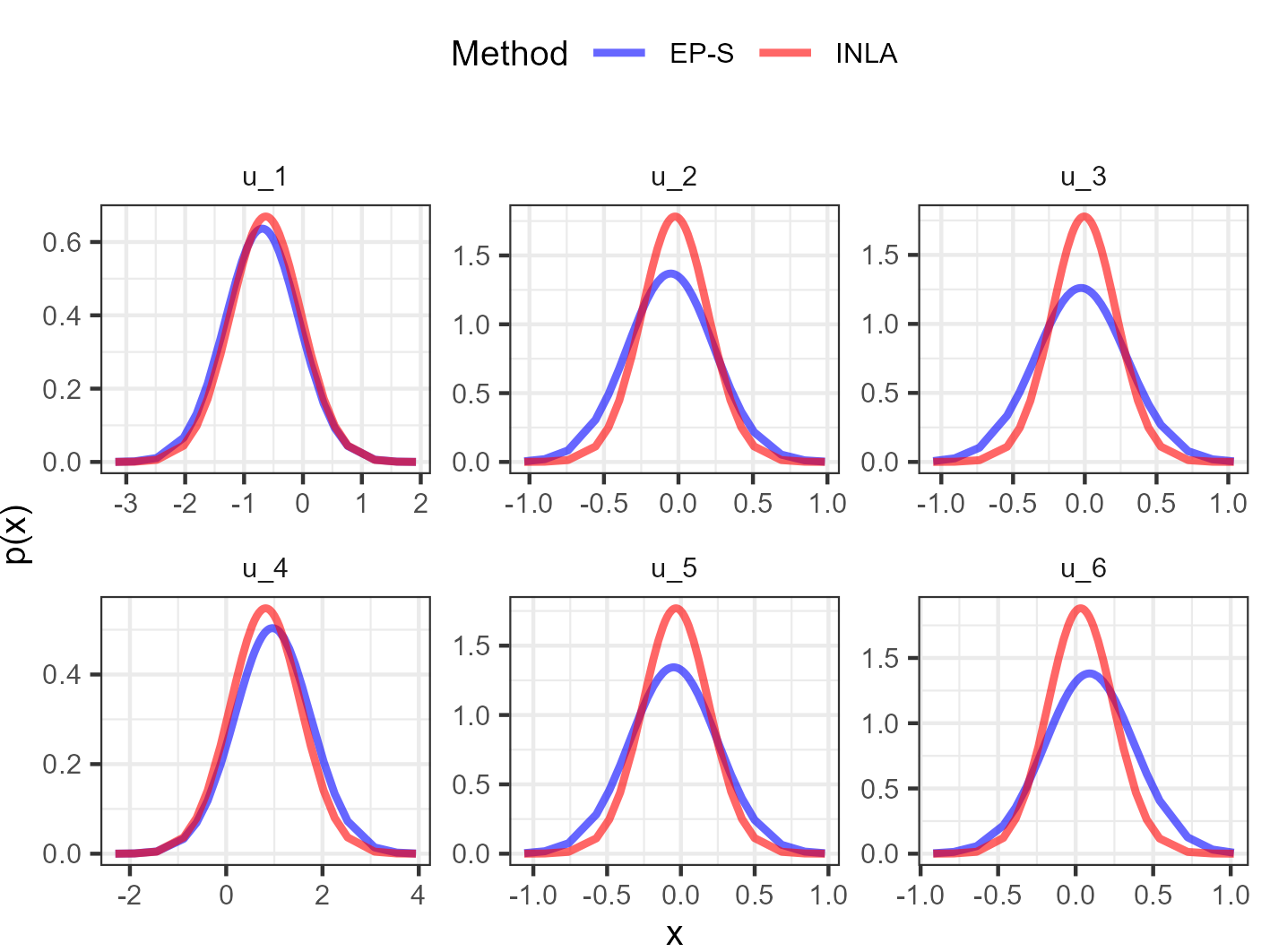}
    \end{center}
    \caption{Bayesian mixed-effects binomial regression model, fit to the NLSY dataset---approximations to marginal posterior distributions for select $\vu$ components.}
    \label{fig:binom_nlsy97_u}
\end{figure}

\begin{figure}
    \begin{center}
        \includegraphics{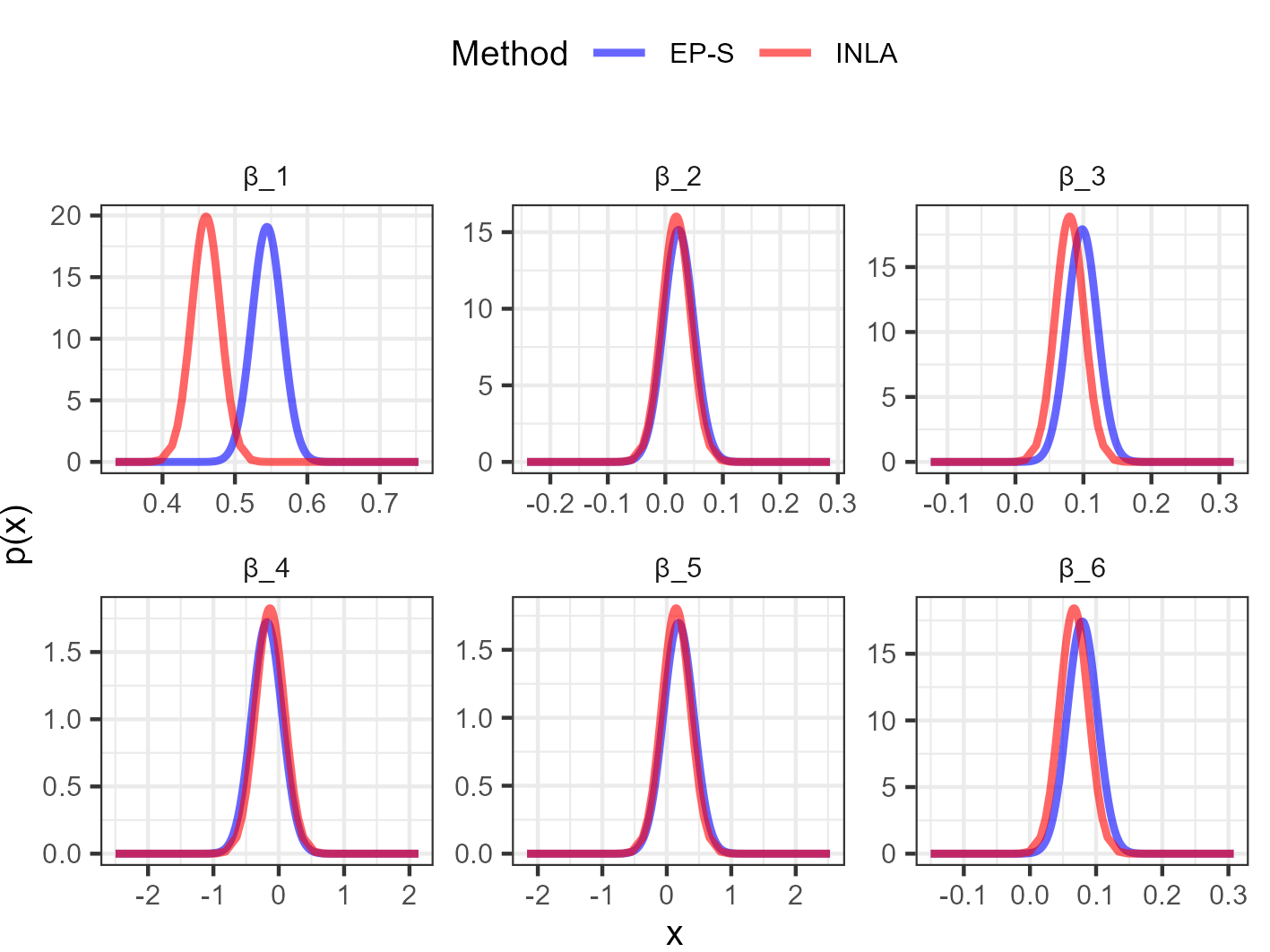}
    \end{center}
    \caption{Bayesian mixed-effects binomial regression model, fit to the NLSY dataset---approximations to marginal posterior distributions for select $\vbeta$ components.}
    \label{fig:binom_nlsy97_beta}
\end{figure}

\begin{figure}
    \begin{center}
        \includegraphics{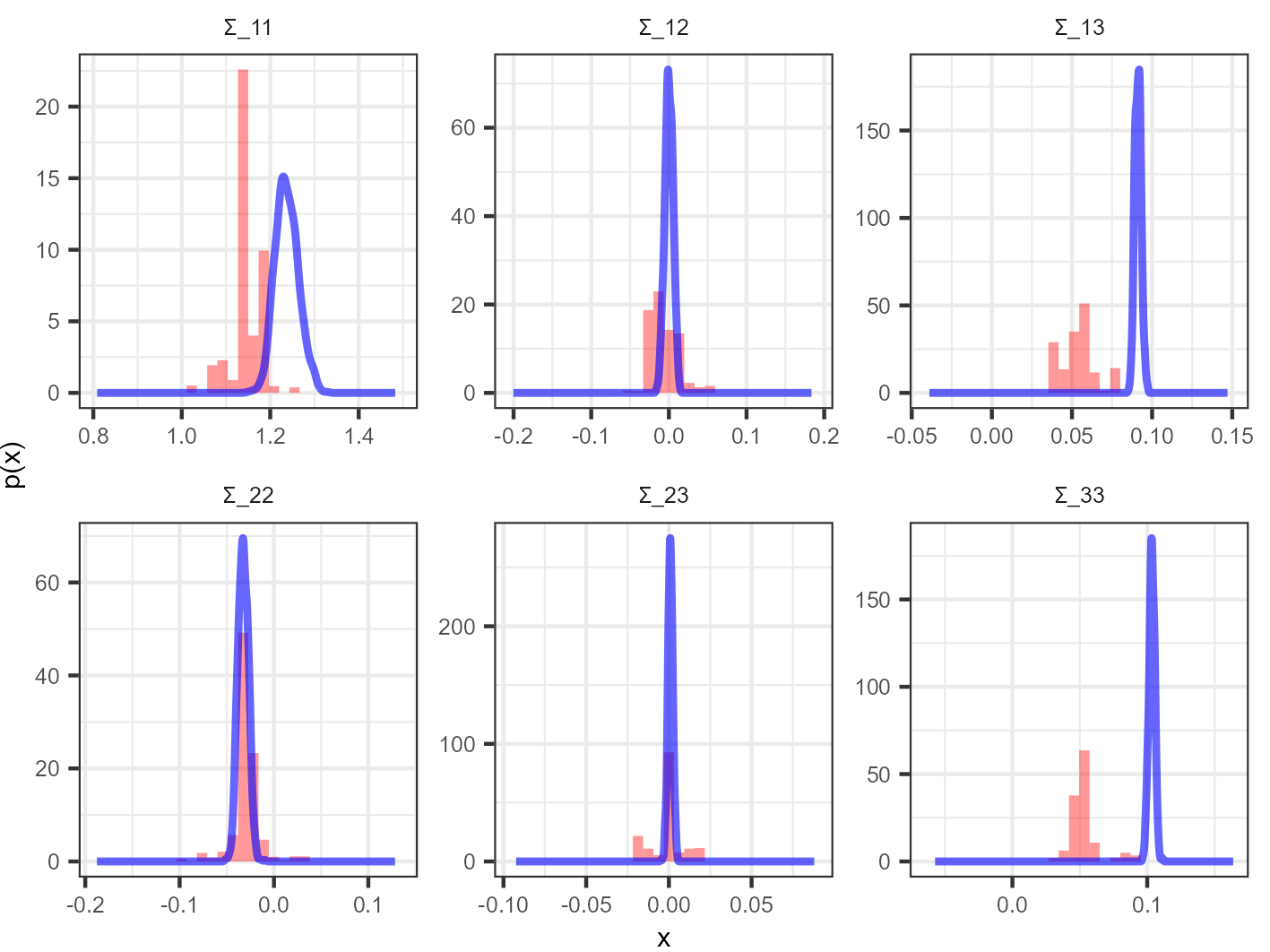}
    \end{center}
    \caption{Bayesian mixed-effects binomial regression model, fit to the NLSY dataset---approximations to marginal posterior distributions for select $\mSigma$ components.}
    \label{fig:binom_nlsy97_Sigma}
\end{figure}

Marginal distribution plots corresponding to the binomial experiments on the NLSY dataset are given from Figure \ref{fig:binom_nlsy97_u} to Figure \ref{fig:binom_nlsy97_Sigma}.
As this is a larger dataset, we now plot the first six marginal distributions of each type.
The plots generally agree with Table \ref{table:binom_marginal_big}.
One exception is that EP-S slightly overestimates the intercept term when compared to INLA.

\subsection{Maximum mean discrepancies}

Throughout the experiments, multivariate accuracy was measured by maximum mean discrepancy (MMD, \citealt{gretton2012kernel}) from the MCMC gold standard.
In particular, we used the unbiased estimate of the squared MMD,
\begin{align*}
    \text{MMD}_u^2=\frac{1}{m(m-1)}\sum_{i\neq j}^m\left[k(\vx^{(i)},\vx^{(j)})+k(\vy^{(i)},\vy^{(j)})-k(\vx^{(i)},\vy^{(j)})-k(\vx^{(j)},\vy^{(i)})\right],
\end{align*}
where lower values indicate higher accuracy.
The kernel function $k(\cdot,\cdot)$ was chosen to be the inner product for simplicity.
The quantity $m$, which we set to 1000, is the total number of samples from each of the two distributions to be compared; the samples are denoted as $\vx^{(i)}$ and $\vy^{(j)}$, and are required to be independent from each other.
In our experiments, we let $\vx^{(i)}$ be the tail samples from the MCMC gold standard, and $\vy^{(j)}$ be the samples from the method to be evaluated.
While it is typical to have $\vx^{(i)}$ and $\vy^{(j)}$ be samples from the joint distribution, we also considered samples from marginal distributions in order to evaluate accuracy when approximating certain types of parameters.
When $\text{MMD}_u^2$ was calculated to be negative, it was set to zero.
Table \ref{table:zip_mmd} and Table \ref{table:binom_mmd} show the MMDs corresponding to the second set of experiments for the ZIP and binomial models respectively.
The rankings of each method are seen to remain mostly unchanged when compared to the marginal accuracies in the main text.

\begin{table}
\centering
\fontsize{9pt}{9pt}\selectfont
\begin{tabular}{@{}cccccccc@{}}
\toprule
\multirow{2}{*}{Dataset}     & \multirow{2}{*}{Method} & \multicolumn{5}{c}{$\text{MMD}^2_u\times10^3$} \\
                             &                         & $(\vtheta,\text{vech}(\mSigma))$       & $\vu$       & $\lambda$      & $\vbeta$      & $\text{vech}(\mSigma)$      &                           \\ \midrule
\multirow{3}{*}{\texttt{Epilepsy}}         & EP-S                      & 45.2 & 23.6 & 21.4 & 0.3 & 0.0               \\
                             & INLA                    & 37.4 & 34.4 &  2.8 & 0.2 & 0.1               \\
                             & GVB                     & 23.9 & 22.6 &  0.6 & 0.5 & 0.3               \\ \midrule
\multirow{3}{*}{\texttt{Fly}} & EP-S                      & 195.5 &  35.5 & 0.0 & 160.0 & 0.0               \\
                             & INLA                    & 122.4 & 122.9 & 0.0 &   0.0 & 0.0               \\
                             & GVB                     & 190.3 &  78.3 & 0.5 & 111.4 & 0.1               \\ \midrule
\multirow{3}{*}{\texttt{Owl\_1}} & EP-S                      & 13.5 & 13.1 & 0.0 & 0.4 & 0.1               \\
                             & INLA                    &  0.8 &  0.7 & 0.1 & 0.0 & 0.1               \\
                             & GVB                     &  8.6 &  8.2 & 0.0 & 0.0 & 0.4               \\ \midrule
\multirow{3}{*}{\texttt{Owl\_3}}         & EP-S                      & 100.6 & 99.1 & 0.1 & 0.1 &   1.3               \\
                             & INLA                    &  22.3 & 22.0 & 0.1 & 0.1 &   0.0               \\
                             & GVB                     & 149.6 & 38.4 & 0.4 & 0.0 & 110.8               \\ \midrule
\multirow{3}{*}{\texttt{Salamanders}} & EP-S                      & 462.8 & 442.1 & 0.1 & 3.7 &  16.9               \\
                             & INLA                    & 129.4 & 124.6 & 0.0 & 1.9 &   2.9               \\
                             & GVB                     & 574.1 & 409.9 & 0.0 & 2.8 & 161.4               \\ \midrule
\multirow{3}{*}{\texttt{Webworms}}    & EP-S                      & 8.1 & 7.4 & 0.1 & 0.6 & 0.0               \\
                             & INLA                    & 0.0 & 0.0 & 0.1 & 0.0 & 0.3               \\
                             & GVB                     & 6.3 & 2.7 & 0.2 & 0.0 & 3.4               \\ \bottomrule
\end{tabular}
\caption{MMDs from MCMC of ABI methods when performing inference on a Bayesian mixed-effects ZIP regression model, for smaller to medium-sized datasets.}
\label{table:zip_mmd}
\end{table}

\begin{table}
\centering
\fontsize{9pt}{9pt}\selectfont
\begin{tabular}{@{}ccccccc@{}}
\toprule
\multirow{2}{*}{Dataset}     & \multirow{2}{*}{Method} & \multicolumn{4}{c}{$\text{MMD}^2_u\times10^3$} \\
                             &                         & $(\vtheta,\text{vech}(\mSigma))$         & $\vu$         & $\vbeta$         & $\text{vech}(\mSigma)$        &                           \\ \midrule
\multirow{3}{*}{\texttt{CTSIB}}         & EP-S                      &  554.8 &  447.6 &  27.6 &  79.6                 \\
                             & INLA                    & 1898.2 & 1574.2 & 297.4 &  26.6                 \\
                             & GVB                     &  529.1 &  317.4 &  11.3 & 200.4                 \\ \midrule
\multirow{3}{*}{\texttt{HDP\_1}} & EP-S                      & 0.9 & 0.9 & 0.0 & 0.0                 \\
                             & INLA                    & 0.7 & 0.7 & 0.0 & 0.0                 \\
                             & GVB                     & 0.0 & 0.0 & 0.0 & 0.1                 \\ \midrule
\multirow{3}{*}{\texttt{HDP\_3}} & EP-S                      & 24.7 & 24.4 & 0.0 &   0.4                 \\
                             & INLA                    &   2.0 &  2.0 & 0.0 &   0.1                 \\
                             & GVB                     & 370.3 & 24.3 & 0.1 & 346.0                 \\ \midrule
\multirow{3}{*}{\texttt{Salamanders}}         & EP-S                      &  10.4 &   9.4 & 0.8 &  0.2                 \\
                             & INLA                    &  42.6 &  41.1 & 0.1 &  1.3                 \\
                             & GVB                     & 185.9 & 122.3 & 0.7 & 62.9                 \\ \midrule
\multirow{3}{*}{\texttt{Tapped}} & EP-S                      &   0.0 &   0.0 &  0.0 &  0.3                 \\
                             & INLA                    & 212.2 & 171.4 & 37.7 &  3.1                 \\
                             & GVB                     & 191.2 & 132.3 &  2.2 & 56.7                 \\ \midrule
\multirow{3}{*}{\texttt{Toenail}}      & EP-S                      &  4659.3 &  4081.3 &  10.7 &  567.3                 \\
                             & INLA                    & 81319.6 & 79464.3 & 322.9 & 1532.4                 \\
                             & GVB                     &  4666.0 &  4264.5 &   0.4 &  401.0                 \\ \bottomrule
\end{tabular}
\caption{MMDs from MCMC of ABI methods when performing inference on a Bayesian mixed-effects binomial regression model, for smaller to medium-sized datasets.}
\label{table:binom_mmd}
\end{table}

Table \ref{table:binom_mmd_big} shows the MMDs corresponding to the binomial experiments on the NLSY dataset.
The large MMD of EP-S from INLA in the $\vu$ component is not surprising, given the large amount of groups and also the large MMDs of INLA from MCMC for the smaller to medium-sized datasets.

\begin{table}
\centering
\fontsize{9pt}{9pt}\selectfont
\begin{tabular}{@{}cccccccc@{}}
\toprule
\multicolumn{4}{c}{$\text{MMD}^2_u\times10^3$} \\
                            $(\vtheta,\text{vech}(\mSigma))$       & $\vu$ & $\vbeta$      & $\text{vech}(\mSigma)$      &                           \\ \midrule
                             79628.1 & 79471.8 & 143.8 & 12.1             \\ \bottomrule
\end{tabular}
\caption{MMDs from INLA of distributed EP-S when performing inference on a Bayesian mixed-effects binomial regression model for the NLSY dataset.}
\label{table:binom_mmd_big}
\end{table}

\bibliographystyle{plainnat}
\bibliography{main.bib}

\end{document}